\documentclass{JHEP3}

\usepackage{amsmath}
\usepackage{amssymb}
\usepackage{mathrsfs}
\usepackage{graphicx}%
\def \eq#1 { \begin{equation} #1 \end{equation} }
\def \eqn#1{ \begin{eqnarray} #1 \end{eqnarray} }
\def \nn { \nonumber }

\def \cHa { \mathcal{H}_{\rm aux} }
\def \cHp { \mathcal{H}_{\rm phys} }

\def\aux{[\!]}                                  % double-line
\def\dket#1{\aux #1 \rangle}                    % double-line ket
\def\dbra#1{\langle #1 \aux}                    % double-line bra
\def\dvac{\aux 0 \rangle}                       % double-line vacuum ket
\def\ket#1{\vert #1 \rangle}                    % ket
\def\brak#1#2{\langle #1 \vert #2 \rangle}      % bracket

\def \half { \frac{1}{2} }

\def \d { \partial }
\def \s { \sigma }
\def \l { \lambda }
\def \wN {\widetilde{N}}
\def \wM {\widetilde{M}}

\def \vk { \vec {k } }
\def \vj { \vec{ j } }
\def \vm { \vec{m} }
\def \vjp { \vec{ j}+1 }
\def \vjm { \vec{ j}-1 }

\def \Td { {}^{(d)}T }

\def \Yd { {}^{(d)}Y }
\def \Ydd { {}^{(d-1)}Y }
\def \cY { \mathcal{ Y } }
\def \cYd { {}^{(d)}\mathcal{ Y } }

\def \chY { \hat{ \mathcal{ Y } } }
\def \Dd { {}^{(d)}D }
\def \Ddd { {}^{(d-1)}D }
\def \Vd { {}^{(d)}V }
\def \Vdd { {}^{(d-1)}V }
\def \Laplaced {\,{}^{(d)}\nabla^2}
\def \Laplacedd {\,{}^{(d-1)}\nabla^2}
\def \LaplaceD {\,{}^{(D)}\nabla^2}
\def \gd {{}^{(d)}\gamma}
\def \gdd {{}^{(d-1)}\gamma}

\def \hB { \hat{B} }

\def \G { \Gamma }
\def \cI { \mathcal{I} }
\def \Lie { \pounds }
\def \Re { \,{\rm Re} }
\def \ad { a^{\dagger} }
\def \Proj { \mathcal{P}_0 }
\def \varphibar { \overline{\varphi} }
\def \sigmabar { \bar{\sigma} }
\def \PG { {\rm P.G. } }

\def \com#1#2{ \left[ #1, #2 \right] }
\def \2F1#1#2#3#4{\,\phantom{}_2F_1\left[#1\,,\,#2\,;\,#3\,;\,#4\right] }
\def \3F2#1#2#3#4#5#6
{\,\phantom{}_3F_2\left[
    \begin{array}{cccc}
      #1\,, & #2\,,  & #3\,; & \\
            &        &       & #6 \\
      #4\,, & #5\,;  &       &
      \end{array} \right]}
\def \KG#1#2{\big\langle #1, #2 \big\rangle_{\rm KG}}

\def \GG#1{\Gamma\left[ #1 \right]}
\def \GGG#1#2{\,\Gamma\left[ \begin{array}{l}
      #1 \\
      #2
    \end{array} \right]}
\def\sech{{\rm sech\,}}
\def\ph{\phantom}

\title{Group Averaging for de Sitter free fields}
\author{Donald Marolf and Ian A. Morrison \\ 
  Department of Physics, University of California at Santa
  Barbara, Santa Barbara, CA 93106, USA \\
  {\it E-mail:} \email{marolf@physics.ucsb.edu}, 
  \email{ian\_morrison@physics.ucsb.edu}
}

\abstract{
  Perturbative gravity about global de Sitter space is subject to
  linearization-stability constraints.  Such constraints imply that quantum
  states of matter fields couple consistently to gravity {\it only} if the
  matter state has vanishing de Sitter charges; i.e., only if the state is
  invariant under the symmetries of de Sitter space.  As noted by Higuchi,
  the usual Fock spaces for matter fields contain no de Sitter-invariant
  states except the vacuum, though a new Hilbert space of de Sitter invariant
  states can be constructed via so-called group-averaging techniques.  We
  study this construction for free scalar fields of arbitrary positive mass
  in any dimension, and for linear vector and tensor gauge fields in any
  dimension. Our main result is to show in each case that group averaging
  converges for states containing a sufficient number of particles.  We
  consider general $N$-particle states with smooth wavefunctions, though
  we obtain somewhat stronger results when the wavefunctions are finite
  linear combinations of de Sitter harmonics.  Along the way we obtain
  explicit expressions for general boost matrix elements in a familiar basis.
}
\preprint{}

\begin{document}
%%%%%%%%%%%%%%%%%%%%%%%%%%%%%%%%%%%%%%%%

%%%%%%%%%%%%%%%%%%%%%%%%%%%%%%%%%%%%%%%%
\section{Introduction} \label{sec:intro}
%%%%%%%%%%%%%%%%%%%%%%%%%%%%%%%%%%%%%%%%

It has been known since the 1970s \cite{BD,VMLS1,VMLS2,Arms,FM,AMM} that perturbative gravity about a background having both compact Cauchy surfaces and Killing symmetries is subject to so-called linearization-stability constraints; i.e., that a linearized perturbation can be extended to a full solution only of it satisfies certain constraints.  These constraints may be summarized by noting that linearized gravity on such backgrounds has conserved charges corresponding to the Killing symmetries.  In the classical context, the constraints state that a solution of linearized gravity can be extended to a solution of the full interacting theory {\it only} if these conserved charges  vanish.  In the quantum context, the vanishing of such charges requires the linearized state to be invariant under the symmetries of the background \cite{VM}.  When matter fields are present, it is the state which describes both matter and linearized gravitational waves propagating on the fixed background which much be invariant.   The physical root of this phenomenon is that Killing symmetries of the background are gauge symmetries when Cauchy surfaces are compact.

Perturbative quantum gravity about de Sitter space presents a particularly interesting case of this phenomenon.  Perturbative states must be invariant under the de Sitter group \cite{AHI,AHII,LU} but, because this group is non-compact, the usual Fock spaces contain no de Sitter invariant states except for a possible vacuum state \cite{AHI}; i.e., they contain far too few states for a useful theory of perturbative gravity.

However, it was shown in \cite{AHII} that one may use a standard Fock space of non-invariant states (which we call the  `auxiliary' space $\cHa$) to construct a new `physical' Hilbert space $\cHp$ of de Sitter invariant states via techniques known as group averaging.  The issues involved are very similar to those that arise in Dirac quantization of constrained systems \cite{Dirac:1964}; indeed, group-averaging was independently introduced in that context \cite{KL,QORD}.  The basic idea is to note that linear combinations of auxiliary states of the form
\eq{
  \label{eq:definitionOfPhysicalStates}
  \ket{\Psi} := \int_{g\in G} dg\, U(g) \dket{\psi}\;
}
are formally de Sitter invariant.  Here $U(g)$ gives the unitary representation of $G$ on $\cHa$ and
$dg$ is the Haar measure on the de Sitter group $G$. Since the de Sitter group is unimodular, $dg$ is invariant under both right- and left-translations.  For compact
groups the integral \eqref{eq:definitionOfPhysicalStates} converges and simply projects the state $\dket{\psi}$ onto the trivial representation of $G$. For non-compact $G$ the state (\ref{eq:definitionOfPhysicalStates}) is not normalizeable in $\cHa$, but can nevertheless often be understood \cite{ALMMT,WhereAreWe} as a well-defined ``generalized state'' in a sense similar to that used in quantum mechanics for the non-normalizeable eigenstates of operators with continuous spectrum (e.g., plane waves in infinite space).  The key questions are whether the so-called group averaged inner product
\eq{
 \label{eq:definitionOfPhysicalInnerProduct}
  \brak{\Psi_1}{\Psi_2} := \dbra{\psi_1} \cdot \ket{\Psi_2}
  = \int_{g\in G} dg\, \dbra{\psi_1} U(g) \dket{\psi_2}\;
}
converges on a sufficiently large set of states, and whether this inner product is positive definite.  If so, it make be taken\footnote{The point of this construction is that, since all gauge-invariant observables commute with $U(g)$, they have a natural action on $\cHp$.  For comments on the extraction of physics from such observables see \cite{VM,QORD,BDT,GMH,GM1}.} as the physical inner product which defines $\cHp$.

The group-averaged inner product (\ref{eq:definitionOfPhysicalInnerProduct}) and the Hilbert space $\cHp$ were studied in detail for the special cases of free 2d massless scalar fields (with the zero-mode removed) and 4d linearized gravitons in \cite{AHII}  and for $4d$ conformally coupled scalars in \cite{AHnotes}.  For these cases it was shown that (\ref{eq:definitionOfPhysicalInnerProduct}) converges when the states $\dket{\psi_{1,2}}$ contain sufficiently many particles and that it is positive definite.  For these special cases it was possible to find a fairly explicit orthonormal basis for $\cHp$.  (See also \cite{Marolf:2008} for a similar treatment including the zero-mode for the 2d massless scalar.)

However, the inner product (\ref{eq:definitionOfPhysicalInnerProduct}) is
more difficult to study for general free fields.  Such an analysis was begun
in \cite{AHnotes}, which showed that group averaging converges for
1+1-dimensional principal series massive scalar fields in certain states
with $N \ge 3$ particles.  Our purpose here is to generalize the known
results to arbitrary dimensions, arbitrary states with smooth $n$-particle wavefunctions, and more general fields.  In particular, we consider arbitrary scalar masses $M^2 > 0$ as well as both vector gauge fields and linearized gravitons.  Our main
result will be that (\ref{eq:definitionOfPhysicalInnerProduct}) converges
whenever the auxiliary states $\dket{\psi_{1,2}}$ contain sufficiently many
particles for either i) general smooth $N$-particle wavefunctions or ii) the special case of wavefunctions given by finite linear combinations of de Sitter Harmonics.
For linear vector gauge fields, linearized gravitons, and scalar fields with masses corresponding to principal series representations of the $dS_{d+1}$ de Sitter group, $d+1$ particles suffice for general smooth wavefunctions.  For scalar fields of smaller mass, we require more particles in a way that will be explained after establishing our conventions below.  For most cases, we are able to show convergence with even fewer particles for the special case (ii).  As discussed in section \ref{disc}, we suspect that these smaller particle numbers may also suffice for the more general case (i).

While our analysis does not guarantee that
(\ref{eq:definitionOfPhysicalInnerProduct}) is positive definite,
our results do show that (\ref{eq:definitionOfPhysicalInnerProduct}) is
positive when $\dket{\psi_{1}}= \dket{\psi_{2}}$ is a tensor product of our de Sitter harmonics.
In addition, a theorem of \cite{MG1} states that when
(\ref{eq:definitionOfPhysicalInnerProduct}) converges it gives the unique
inner product on states of the form \eqref{eq:definitionOfPhysicalStates}
consistent with the *-algebra of bounded gauge-invariant observables on
$\cHa$.  Thus on physical grounds one expects that
(\ref{eq:definitionOfPhysicalInnerProduct}) is positive
definite\footnote{See \cite{JLAM03,JLSum,JLAM05} for further comments on the
positivity of (\ref{eq:definitionOfPhysicalInnerProduct}) in other settings.}.

Our discussion is organized as follows.
Section \ref{sec:scalar} examines group averaging for massive scalar
fields and  obtains explicit expressions for the matrix elements of $U(g)$
in a familiar basis, though certain technical details are relegated to the
appendices. Such expressions allow us to study the convergence of
(\ref{eq:definitionOfPhysicalInnerProduct}) for scalar fields, and to
establish that group averaging converges under the conditions stated above.
We then perform a similar study for vector and tensor gauge fields in
section \ref{HS}.  We find that many results from the scalar case can
be applied directly to the vector and tensor cases.
We conclude with a brief discussion of open issues and states with small particle numbers in section \ref{disc}.

%%%%%%%%%%%%%%%%%%%%%%%%%%%%%%%%%%%%%%%%
\section{Massive scalar fields} \label{sec:scalar}
%%%%%%%%%%%%%%%%%%%%%%%%%%%%%%%%%%%%%%%%

We wish to analyze the convergence of group averaging for
massive scalar fields in $D = d+1$ dimensional de Sitter space, generalizing
 and extending certain results of \cite{AHnotes} for the case $D=2$.
Our main task will be to construct an explicit expression for the matrix
elements $ \dbra{\psi_1} B(\lambda) \dket{\psi_2}$ for states
$\dket{\psi_{1,2}}$ in the standard de Sitter Fock space of a free massive
scalar field, from which the convergence properties of
(\ref{eq:definitionOfPhysicalInnerProduct}) can be easily evaluated.
Here $B(\lambda)$ is a finite boost of rapidity $\lambda$. As discussed
in section \ref{convergence} below, such matrix elements control the convergence
of (\ref{eq:definitionOfPhysicalInnerProduct}).

In an $N$-particle state, boost matrix elements reduce to (symmetrized)
products of matrix elements for one-particle states.  We will therefore
focus on computing $ \dbra{\psi_1} B(\lambda) \dket{\psi_2}$ for one-particle
states below. These are just  matrix elements for irreducible representations
of the de Sitter group.  There is a long history of such computations in the
literature, beginning with Bargmann's work \cite{Bargmann:1947} on SO(2,1)
and continued in  \cite{Dixmier:1960,Takahashi} for SO(3,1); see also
\cite{Varlamov:2002dr,Varlamov:2006ce}.   These works compute matrix
elements in the basis we use below.   Using the rather different
Gelfand-Tsetlin basis \cite{GT}, a general formula for all SO($D$,1) was
found in \cite{Wolf:1971mv} and was evaluated explicitly in
\cite{Wong:1981dd} in terms of hypergeometric functions. Appendix
\ref{app:integral} is devoted to providing a more direct computation of
the general SO($D$,1) matrix elements for scalar representations in a
basis more familiar to physicists.  For our purposes the results of
appendix \ref{app:integral} are somewhat simpler than those of
\cite{Wong:1981dd} and give us better control over the asymptotics at
large $g$.

We establish notation and review scalar fields on de Sitter in section
\ref{scalarRev}.  Section \ref{BME} computes the desired matrix elements
of $B(\lambda)$ using results from the appendices.  The convergence of
(\ref{eq:definitionOfPhysicalInnerProduct}) is then analyzed in section
\ref{convergence}.

%%%%%%%%%%%%%%%%%%%%%%%%%%%%%%%%%%%%%%%%
\subsection{Scalar Fields in de Sitter}
%%%%%%%%%%%%%%%%%%%%%%%%%%%%%%%%%%%%%%%%
\label{scalarRev}

Recall that the de Sitter metric may be conveniently described by embedding
$dS_D$ in a $D+1$-dimensional Minkowski space $M_{D+1}$.  Using Cartesian coordinates $X^A$ in $M_{D+1}$, one may define de Sitter spacetime as the hyperboloid of constant radius $\ell$:
\eq{
  \ell^2 = \eta_{AB}X^A X^B
  = - (X^0)^2 + (X^1)^2 + \dots + (X^D)^2 \; ,
}
where $\eta_{AB}$ is the $D+1$ Minkowski metric.
The induced metric can be written
\eqn{ \label{eq:dSMetric}
  ds^2 &=& \ell^2 \left[ - dt^2 + (\cosh^2t) d\Omega_d^2 \right] \nn \\
  &=& \ell^2\left[-dt^2 + (\cosh^2 t)
  \left( d\chi^2 + \sin^2\chi \, d\Omega_{d-1}^2 \right)\right]\; ,
}
where $t = \sinh^{-1}(X^0/\ell)$, $d = D - 1$, and $d\Omega_d^2$ is
the metric of the $d$-sphere $S^d$.

We consider scalar fields satisfying the Klein-Gordon equation
\eq{ \label{eq:scalarEOM}
  \Box \varphi =  \left(M^2 + \xi R \right) \varphi
  =: - \left[ \frac{\sigma(\sigma+d)}{\ell^2} \right] \varphi\;,
}
where $M^2$ is the mass, $R = D(D-1)/\ell^2$ is the scalar curvature
of de Sitter, $\xi$ is a non-minimal coupling constant, and
$\Box := g^{\mu\nu}\nabla_\mu\nabla_\nu$ is the wave operator
on de Sitter. We define the complex parameter $\sigma$ through the
last equality. Below, we focus on the case $\sigma \neq 0, -d$, though this case can be included in the analysis below if the zero-mode is removed.  (See \cite{Marolf:2008} for a discussion of group-averaging the $\sigma=0$ zero-mode for $D=2$.)

One may quantize the scalar field $\varphi_\sigma$ with Casimir $-\sigma(\sigma+d)$ by expanding it in a sum over modes
\eq{
  \varphi_{\sigma}(x) =
  \sum_{\vj} a_{\vj} \varphi_{\sigma \vj}(x) + \ad_{\vj} \varphibar_{\sigma\vj}(x)\;,
}
where the $\varphi_{\sigma \vj}$ form an irreducible representation of the de Sitter group. The vector ${\vj} =  j_d, \dots, j_1$ contains $d$ angular momentum quantum numbers which satisfy
\eq{
  \label{eq:js}
  j_d \ge j_{d-1} \ge \dots \ge |j_1| \;; .
}
Below, it will be convenient to denote the total angular momentum by $j := j_d$ and the ``second''
quantum number by $k := j_{d-1}$.  See \cite{Allen:1985ux,Allen:1987tz,Mottola:1984ar,Polarski:1990ux} for classic references on quantizing such fields.

We take our modes $\varphi_{\sigma {\vj}}$ to be normalized with respect to the Klein-Gordon inner product
\eq{
\label{sKGnorm}
  \KG{\varphi_{\sigma \vj}}{\varphi_{\sigma \vm}}
 = -i
  \int d\Sigma^\mu \, \varphi_{\sigma \vj} \overleftrightarrow{\nabla_\mu}
  \overline{\varphi}_{\sigma \vm}
  = \delta_{\vj\vm}\;,
}
so that the creation and annihilation operators satisfy the usual algebra:
\eq{
\label{comaa}
  \com{a_{\vj}}{\ad_{\vm}} = \delta_{\vj \vm}\;, \quad
  \com{a_{\vj}}{a_{\vm}} = \com{\ad_{\vj}}{\ad_{\vm}} = 0\;.
}
They can also be shown to satisfy a de Sitter invariant positive frequency condition (see e.g. \cite{AF}).

The auxiliary Hilbert space $\cHa$ is then built from the
vacuum $\dvac$ satisfying
\eq{
  a_{\vj} \dvac = 0\quad \forall\, \vj\;.
}
The discussion above is explicitly de Sitter-invariant, so $\dvac$ is the familiar de Sitter-invariant (Bunch-Davies) vacuum.  The full auxiliary space $\cHa$ is spanned by states of the form
$\dket{\psi} = \ad_{\vj_1} \dots \ad_{\vj_n} \dvac$, whose inner products are readily computed from  (\ref{sKGnorm}) and (\ref{comaa}).

The one-particle states of $\cHa$ form a scalar representation of $SO_0(D,1)$.  Recall \cite{Vilenkin:1991,Bargmann:1947,Wong:1974cv,Schwarz:1971,Ottoson:1967}
that scalar representations $T^{D\sigma}$ of $SO_0(D,1)$
are labeled by the eigenvalue of the quadratic Casimir, whose action on one-particle states is given by the de Sitter  Laplace-Beltrami operator.  Thus, one-particle states satisfying
(\ref{eq:scalarEOM}) have Casimir eigenvalues $-\sigma(\sigma+d)$.
There is a redundancy in this description, as (\ref{eq:scalarEOM}) is unchanged under the interchange
\eq{ \label{eq:sigmaRedundancy}
  \sigma \to -(\sigma + d)\;.
}
Unitary representations must have real Casimirs which requires
\eq{
  \sigma = \sigmabar, \ \ {\rm or} \
 \ \quad \sigma + \sigmabar = - d\;, .
}
This gives three distinct series of representations
\cite{Vilenkin:1991,Wong:1974cv}
\begin{enumerate}

\item
  Principal series: representations with $\sigma = -\frac{d}{2} + i \rho$,
  $\rho \in \mathbb{R}$.

\item
  Complementary series: representations with $-d < \sigma < 0$,
  $\sigma \in \mathbb{R}$.

\item
  Discrete series: representations with $\sigma > - \frac{d}{2}$,
  $\sigma \in \mathbb{Z}$.

\end{enumerate}
If we restrict attention to the non-tachyonic cases
$\ell^2(M^2 +\xi R) = - \sigma(\sigma+d) \ge 0$ and use
(\ref{eq:sigmaRedundancy}) we can restrict the range of
$\sigma$ to
\eq{ \label{eq:sigmaRange}
  -\frac{d}{2} \le \sigma \le 0, \quad
  \sigma = - \frac{d}{2} + i \rho, \; \rho \in \mathbb{R}\;.
}
In particular, the ``massless'' scalar field with $\ell^2(M^2 + \xi R) = 0$
corresponds to $\sigma = 0$; increasing $\ell^2(M^2 + \xi R)$ moves $\sigma$
along the real line from $0$ to $-d/2$; further increasing $\ell^2(M^2 + \xi R)$
moves $\sigma$ off the real line along $\sigma = -\frac{d}{2} + i \rho$.

Finally, let us return to the equation of motion (\ref{eq:scalarEOM}) and construct an explicit set of mode functions $\varphi_{\sigma {\vj}}$ for each $\sigma$ in the above range. To
do so,
let $\Laplaced$ be the Laplacian on $S^d$ and note that the rescaled wave operator
\eq{ \label{eq:scalarBox}
\ell^2   \Box
  = \left[-\d_t^2 - d (\tanh t) \, \d_t + \cosh^2 t \Laplaced\right]\;
}
is the analytic continuation to $dS_D$ of the Laplacian $\LaplaceD$ on $S^D$.     As a result, solutions
to (\ref{eq:scalarEOM}) are analytically continued scalar spherical harmonics.   Such spherical harmonics are reviewed in
Appendix~\ref{app:scalar} and lead to a basis of solutions on $dS_D$ of the form
\eq{ \label{eq:scalarModes}
  \varphi_{\sigma \vj}(t, \Omega_d) = \ell^{(1-d)/2}
  \chY_{\sigma j}(t)\, \Yd_{\vj}(\Omega_d) \; .
}
Here $\Yd_{\vj}(\Omega_d)$ are scalar spherical harmonics on the spatial
$S^d$ slices, while the time-dependent factor in (\ref{eq:scalarModes})
is given by \cite{Higuchi:1986wu}
\eqn{ \label{eq:chY}
  \chY_{\sigma j}(t) &=& N_{d\sigma j}
  (\cosh t)^{-(d-1)/2} P^{-(j+(d-1)/2)}_{\sigma+(d-1)/2}(i \sinh t) \;.
}
In (\ref{eq:chY}), $P^{-\mu}_\nu(x)$ is the Legendre function \cite{GR} and
$N_{d\sigma j}$ is a normalization constant
\eq{
  N_{d\sigma j}
  = \Big[ \frac{\G(j-\sigma)\G(j+\sigma+d)}{2}\Big]^{1/2}\;.
}
Note that $N_{d\sigma j}$ is real for all values of
$\sigma$ (\ref{eq:sigmaRange}).

%%%%%%%%%%%%%%%%%%%%%%%%%%%%%%%%%%%%%%%%
\subsection{Boost matrix elements}
%%%%%%%%%%%%%%%%%%%%%%%%%%%%%%%%%%%%%%%%
\label{BME}

We are now ready to compute matrix elements of the finite boost $B(\lambda)$.  In order to compare with the vector and tensor cases in section \ref{HS}, it is useful to begin with the infinitesimal generator $\hat B$.  To this end, consider the de Sitter Killing field
\eq{ \label{eq:boostVector}
  B^\mu = (\cos\chi, - \tanh t\, \sin\chi, \vec{0}) \;.
}
The corresponding boost generator $\hB$ acts on $\varphi_\sigma$ as
\eq{
  \com{\hB}{\varphi_\sigma} = i \Lie_B \varphi_\sigma
  = i \left( \cos\chi \d_t - \tanh t \sin \chi \d_\chi \right) \varphi_\sigma\;.
}
Using the well-known properties of Legendre functions \cite{GR}
one finds
\eq{\label{boost1}
i \pounds_B \varphi_{\sigma \vj}
  = b^{+}_{\sigma\vj}\, \varphi_{\sigma \vjp}
  + b^{-}_{\sigma\vj}\, \varphi_{\sigma \vjm} \;,
}
where ${\vjp}$ and ${\vjm}$ are $\vj$ with $j$ replaced by $j \pm 1$ and all other $j_m$ unchanged.  The coefficients $b^\pm_{\sigma\vj}$ are
\eqn{
  \label{eq:bpm}
  b^+_{\sigma\vj} &=& \left[
    \frac{(j+k+d-1)(j-k+1)(j-\sigma)(j+\sigma+d)}
    {(2j+d-1)(2j+d+1)} \right]^{1/2}, \nn \\
  b^-_{\sigma\vj} &=& \left[
    \frac{(j+k+d-2)(j-k)(j-1-\sigma)(j-1+\sigma+d)}
    {(2j+d-1)(2j+d-3)} \right]^{1/2} \;.
}
Note that $b^-_{\sigma\vj} =0$ when $j=k$ so that the coefficient of
$\varphi_{\sigma \vjm}$ vanishes when $\varphi_{\sigma \vjm}$ is not defined.  As expected, the boost generator
\eq{
  \hB = - \sum_{\vj} \ad_{\vj}
  (b^-_{\sigma\vj} \,a_{\vjm} + b^+_{\sigma\vj} \,a_{\vjp}) \; ,
}
is Hermitian.

Let $B(\lambda) = e^{i \lambda \hat B}$ be the operator which translates the field
by an amount $\lambda$ along (\ref{eq:boostVector}).  $B(\lambda)$ acts on
scalar fields via the coordinate change $x \to x'$ given by
\eqn{ \label{eq:coordinateChange}
  \sinh t'
  &=& \cosh\lambda \sinh t + \sinh\lambda \cosh t \cos\chi\;, \nn \\
  \cosh t' \cos \chi'
  &=& \sinh \lambda \sinh t + \cosh \lambda \cosh t \cos\chi\;, \nn \\
  \cosh t' \sin\chi'
  &=& \cosh t \sin\chi\;.
}
We wish to compute the matrix elements
\eq{ \label{eq:definitionOfFiniteMatrix}
  B_{\vj \vm}(\lambda)
  := \KG{ B(\lambda) \varphi_{\sigma \vj}}{ \varphi_{\sigma \vm} }
}
of $B(\lambda)$ between one-particle states associated with the mode functions (\ref{eq:scalarModes}). Due to rotational symmetry about the boost axis, the matrix elements $B_{\vj \vm}(\lambda)$ are diagonal
in all angular momenta except the total angular momenta $j,m$; i.e.
$B_{\vj \vm}(\lambda) \propto \delta_{j_1 m_1}\cdots\delta_{j_{d-1},m_{d-1}}$.
Furthermore, the matrix elements are symmetric in $j \leftrightarrow m$ and depend only on $j$, $m$, and the (mutual) second quantum number $k = j_{d-1} = m_{d-1}$. See \cite{Vilenkin:1991} for a group theoretic
explanation of these facts.
As a result, we may adopt a simpler notation for the matrix elements, denoting them by $B_{jmk}(\lambda) = B_{mjk}(\lambda)$.

Using the explicit form (\ref{eq:scalarModes}) of the modes and
the coordinate change (\ref{eq:coordinateChange}), one can compute
the inner product (\ref{eq:definitionOfFiniteMatrix})
directly. This lengthy calculation is performed in
appendix~\ref{app:integral} and yields
\eqn{ \label{eq:solution}
  B_{j m k}(\lambda)
  &=& \mathcal{A}^\sigma_{j m k} \,(\cosh \lambda)^{j-k}\, (i\sinh \lambda)^f
  \Bigg\{
    \sum_{a=0}^{[(m-k)/2]}\sum_{b=0}^{[(j-k)/2]} \mathcal{B}^{ab}_{jmk}
  (\cosh \lambda)^{-2b} \nn \\
  & & \times
  \3F2{\frac{j-\sigma+f}{2}}{\frac{j+\sigma+d+f}{2}}{\frac{j+m+f+1}{2}-k-a-b}
  {\half+f}{\frac{j+m+d+1+f}{2}}{-\sinh^2 \lambda} \Bigg\}.
}
Here ${}_3F_2(a,b,c;d,e;z)$ is the generalized hypergeometric function
(\cite{Bateman,Slater})
the coefficients $\mathcal{A}^\sigma_{j m k}$ and $\mathcal{B}_{j m k}^{ab}$
are defined in (\ref{eq:curlyADef}) and (\ref{eq:curlyBDef}), and $f$
is defined below (\ref{eq:I2}).
We conclude this subsection some comments on this result:

\begin{itemize}

\item{} Due to some asymmetries in our treatment of the $\vj,\vm$ mode
functions, the result (\ref{eq:solution}) is not manifestly symmetric in $j$
and $m$.  However, for any given example one can readily verify that (\ref{eq:solution}) is symmetric by, e.g., using Mathematica to expand (\ref{eq:solution}) as a power series in $\sinh \lambda$.

\item{} It is well known that the boost matrix elements satisfy a number
of recursion relations \cite{Bargmann:1947,Barut:1976}. While we
will not use any such relations here, we note one of the
most useful relations here in order to connect with the work
of previous authors,
\eq{ \label{eq:recursionRelation2}
  b_{\sigma\vj}^+\, B_{j+1,m,k}(\lambda)
  + i \d_\lambda B_{j m k}(\lambda)
  + b_{\sigma\vj}^- \,B_{j-1,m,k}(\lambda) = 0 \;.
}

\item{}
The result (\ref{eq:solution}) is rather cumbersome in general,
but simplifies greatly for two cases. The first is the case where the
scalar field is conformally coupled, $\sigma = -(d-1)/2$, for which
(\ref{eq:solution}) reduces to
\eqn{
  B_{j m k}(\lambda)
  &=& 2^{j-m} \beta_{j m k}\left(\sigma = -\frac{d-1}{2}\right)
  \left(\sinh \frac{\lambda}{2}\right)^{j-m}
  \left(\cosh \frac{\lambda}{2}\right)^{j+m+d-1} \nn \\ & & \times
  \2F1{j+k+d-1}{j-k+1}{j-m+1}{-\sinh^2 \frac{\lambda}{2}} ,
  \label{eq:conformal}
}
where in general we define
\eqn{
  \label{eq:beta}
  \beta_{j m k}(\sigma) &=& \frac{i^{j-m}}{2^{j-m}\G(1+j-m)}
  \nn \\ & & \times
  \Bigg\{
  \GGG{ j+k+d-1, j-k+1, j-\s, j+\s+d, m+\frac{d-1}{2}, m+\frac{d+1}{2} }
  { m+k+d-1, m-k+1, m-\s, m+\s+d, j+\frac{d-1}{2}, j+\frac{d+1}{2} }
  \Bigg\}^{1/2}\;.
  \nn \\
}

Here $\G[\dots]$ denotes a product (and quotient) of gamma functions;
e.g $\GGG{ab}{c} = \frac{\Gamma[a] \Gamma[b]}{\Gamma[c]}$.
Both (\ref{eq:conformal}) and (\ref{eq:beta}) assume $j \ge m$.
A second case in which the matrix elements simplify is when $m=k$.
In this case (\ref{eq:solution}) becomes
\eqn{
  \label{eq:Bjkk}
  B_{j k k}(\lambda) &=& B_{k j k}(\lambda) \nn \\
  &=& \beta_{j k k}(\sigma) (\sinh \lambda)^{j-k}
  \2F1{\frac{j-\sigma}{2}}{\frac{j+\sigma+d}{2}}{j+\frac{d+1}{2}}
  {-\sinh^2 \lambda} \nn \\
  &=& 2^{j+(d-1)/2}\G\left(j+\frac{d+1}{2}\right) \beta_{j k k}(\sigma)
  (\sinh \lambda)^{-k-(d-1)/2}
  P_{\sigma +(d-1)/2}^{-(j+(d-1)/2)}(\cosh \lambda) . \nn \\
}
Matrix elements for nearby $B_{j, k+n,k}(\lambda)$ can then be found
by successive use of the recursion relation (\ref{eq:recursionRelation2}).

\end{itemize}

%%%%%%%%%%%%%%%%%%%%%%%%%%%%%%%%%%%%%%%%
\subsection{Convergence of group averaging}\label{convergence}
%%%%%%%%%%%%%%%%%%%%%%%%%%%%%%%%%%%%%%%%

We wish to use group averaging over the de Sitter group $G = SO_0(D,1)$
to construct physical states which satisfy $M_{AB} \ket{\Psi } = 0$, where
$M_{AB}$ are the generators of the group.
It is useful to first find simple expressions for unitary transformations
$U(g)$ and the Haar measure $dg$ of the de Sitter group.
Under the Cartan decomposition of $SO_0(D,1)$ we can decompose any
group element into two $SO(D)$ rotations and a boost \cite{Vilenkin:1991}:
\eq{
  U(g) = U(\alpha) B(\lambda) U(\gamma)\;,
}
where $\alpha$ and $\gamma$ are group elements of $SO(D) \subset SO(D,1)$ and $B(\lambda)$ is the unitary action in the given representation of a boost of rapidity $\lambda$ along some given Cartesian axis in $M_{D+1}$.  We take the coordinate $\chi$ to have values  $\chi = 0, \pi$ on this axis.
In a similar fashion, the Haar measure can be decomposed as
\cite{Gomberoff:1998ms}
\eq{
  dg = d\alpha\  d\gamma \ d\lambda \  (\sinh\lambda)^d\;,
}
where again $d = D-1$ and $d\alpha, d\gamma$ are both the Haar measure on $SO(D)$. The group averaging inner product
(\ref{eq:definitionOfPhysicalInnerProduct}) is then
\eqn{
\label{simpleGAIP}
  \brak{\Psi_1}{\Psi_2}
  &=&
  \int d\alpha\,d\gamma\,d\lambda\,(\sinh\lambda)^d\,
  \dbra{\psi_1} U(\alpha) B(\lambda) U(\gamma) \dket{\psi_2} \nn\\
  &=&
  \int_0^\infty d\lambda\,(\sinh\lambda)^d
  \dbra{\psi_1} \Proj B(\lambda) \Proj \dket{\psi_2}\;,
}
where in the second line we used the fact that the projector $\Proj$ onto SO(D)-invariant states satisfies
\eqn{\label{eq:definitionOfProjectionOp}
\Proj  = \int d \alpha \ U(\alpha).
}

Due to the projectors $\Proj$, it is only necessary to compute  (\ref{simpleGAIP}) for rotationally invariant auxiliary states $\dket{\psi_{1,2}}$, in which case the action of the projectors is trivial.  In practice, it will often be useful to simply delete the projectors and to consider the integral
\eqn{
  \label{noproj}
 \cI = \int_0^\infty d\lambda\,(\sinh\lambda)^d
  \dbra{\psi_1} B(\lambda) \dket{\psi_2}\; .
 }
Because $SO(D)$ is a compact group, the operators $\Proj$
do not affect the convergence properties of the group averaging
inner product.  Examining the convergence properties of $\cI$
is thus equivalent to examining those of $\brak{\Psi_1}{\Psi_2}$.

For  $N$-particle states in which each particle occupies a definite mode, the integral $\cI$ may be written
\eqn{
  \cI &=&
  \int_0^\infty d\lambda (\sinh\lambda)^d
  \dbra{\vj_1,\dots,\vj_N}
  B(\lambda)
  \dket{\vm_1,\dots,\vm_N}\nn\\
  &=&
  \int_0^\infty d\lambda (\sinh\lambda)^d
  B_{{\vj_1},  ( {\vm_1}}(\lambda)\cdots B_{|{\vj_N}|, \vm_N)}(\lambda) \;,
  \label{eq:NParticleI}
}
where the product of matrix elements $ B_{\vj_1,\vm_1} \cdots B_{\vj_N,\vm_N} $ in
the last line has been symmetrized over the indices ${\vm_1} \cdots {\vm_N}$.
For large $\lambda$ the boost matrix elements have the form
\eqn{
  B_{jmk}(\lambda) &=& \mathcal{C}_{0\,jmk} (e^\lambda)^{\sigma-k}
  \left[ 1 + \mathcal{C}_{1\,j m k} (e^{-2\lambda}) + O(e^{-4\lambda}) \right]
  \nn \\ & &
  + \mathcal{D}_{0\,jmk} (e^\lambda)^{-d-\sigma-k}
  \left[ 1 + \mathcal{D}_{1\,j m k} (e^{-2\lambda}) + O(e^{-4\lambda}) \right]
  \label{eq:BAssymptotics}
}
where $\mathcal{C}_i$ are coefficients that do not depend on $\lambda$
and $\mathcal{D}_i = \mathcal{C}_i(\sigma \to -(\sigma +d ))$.
Therefore, the least convergent part of the integrand in
(\ref{eq:NParticleI}) has the asymptotic behavior
$\sim \exp[\lambda(d+N \sigma -\sum_{i} k_i)]$; the integral converges
for
\eq{
\label{scalarConv}
  \Re\left( d + N \sigma - \sum_{i=1}^N k_i \right) < 0 \;.
}
This has the following results for scalar representations:
\begin{itemize}

\item{}
{\bf Principal series:} The integral
(\ref{eq:NParticleI}) converges absolutely for $N > N_{\rm conv} = 2$.

\item{}
{\bf Complementary and Discrete series with $\sigma \neq 0$:} The integral (\ref{eq:NParticleI})
converges absolutely for $N > N_{\rm conv} = -d/\sigma$. This can in
principle be a very large number for $\sigma$ near 0.

\item{}
{\bf Massless scalars:}
The matrix elements $B_{j m k}(\lambda)$ above do indeed form a representation
of the de Sitter group for $\sigma = 0$. However, this representation does not include the full physics of the zero mode as this representation describes only states with vanishing zero-mode momentum;
see \cite{Marolf:2008}.

\end{itemize}
In all cases the integral may converge for $N < N_{\rm conv}$ for
sufficiently large momenta $k_i$.

%%%%%%%%%%%%%%%%%%%%%%%%%%%%%%%%%%%%%%%%%%%%%%%%%%%%%%%%%%%%%%%%%
\subsection{Convergence for more general states} \label{sec:bound}
%%%%%%%%%%%%%%%%%%%%%%%%%%%%%%%%%%%%%%%%%%%%%%%%%%%%%%%%%%%%%%%%%

The convergence criteria listed above have been derived for states
in which each particle is described by a single mode in our basis.
One would also like to examine more general states which allow the
particles to be in a superposition of modes. Finite superpositions
clearly do not alter the above criteria. However,
understanding the convergence of infinite superpositions requires
some control over the group averaging norms of the states
$\dket{\vj_1,\dots,\vj_N} $ as a function of the angular momenta.
To see this, let us consider an arbitrary smooth $N$-particle
state $\dket{\psi_N}$ which may written as an infinite sum
\eq{ \label{eq:smooth}
  \dket{\psi_N} = \sum_{\vj_1,\dots,\vj_N} c_{\vj_1,\dots,\vj_N}
  \dket{\vj_1, \dots, \vj_n}
}
with coefficients $c_{\vj_1,\dots,\vj_N}$ which decay with all $\vj_i$ faster
than any polynomial. The group averaging inner product of two such
smooth states will be finite so long as the inner product of the
basis states grows no faster than a polynomial in momenta $\vj_i$.

In principle one can compute the group averaging inner product
(\ref{eq:definitionOfPhysicalInnerProduct}) exactly
using the finite boost matrix elements (\ref{eq:solution}) and
$SO(D)$ Clebsch-Gordan coefficients. Unfortunately, it is rather difficult
to extract the angular momentum dependence of this result; indeed,
even if one omits the $SO(D)$ Clebsch-Gordan coefficients as in the previous
section, it is difficult to extract the angular momentum dependence of the
finite boost matrix elements $B_{j m k}(\l)$. In order to proceed,
in this section we derive a simple bound on the
finite boost matrix elements and use this bound to investigate the
dependence of the group averaging inner product on angular momenta $j,m,k$.
The bound's benefit is that it's dependence on angular
momenta is clear; it's weakness is that it has a weaker
fall-off with respect to rapidity $\lambda$ than the actual matrix elements.
As a result, larger numbers of particles will be needed to guarantee
convergence for general smooth wavefunctions.

The calculation of the bound on $B_{j m k}(\l)$ will be somewhat technical.
We therefore begin by summarizing our results, and then follow with
the derivation. We are able to show that the matrix elements
$|B_{j m k}(\l)| \le f(j,m,k) g(\l)$, where $f(j,m,k)$ contains all the
angular momenta dependence and $g(\l)$ contains all the rapidity dependence.
The function $f(j,m,k)$ grows at large momenta no faster than a polynomial
of order $\mathcal{O}( (jm)^{(d+1)/2}k^{-d} )$. The function $g(\l)$
has asymptotic behavior at large $\l \gg 1$ given by

\begin{equation} \label{g}
g(\lambda) \sim \Bigg \{
{ {e^{- \lambda} \  \ \ {\rm for} \  (k - \Re \ \sigma) > 1}
\atop {e^{- (k - \Re \ \sigma) \lambda} \  \ \ {\rm for} \  (k - \Re \ \sigma) \le 1}}
\end{equation}
Recall that
the exact expression for the matrix elements decays as
$\sim \exp[\l (\s-k)]$. Bounding $B_{j m k}(\l)$ in this way thus looses the
additional suppression that occurs for large mass and large angular
momenta $k$. However, it allows us to conclude the following for arbitrary smooth states:

\begin{itemize}

\item{}
  {\bf Principal series:}
  For $d \ge 2$ principal series representations always satisfy
  $\Re\,\s \le -1$.
  Thus for this series our bound on the matrix elements decays with
  $\l$ like $\sim e^{-\l}$. It follows that for $N >  d$
  group averaging converges for any smooth states.

\item{}
  {\bf Complementary series with $\s \le -1$ or $k > 0$:}
  For $d > 2$ the complementary series contains representations with
  $\s \le -1$. The conclusion is as for the principal series:
  for $N >  d$ group averaging converges for any smooth states.

\item{}
  {\bf Complementary series with $\s > -1$ and $k = 0$:}
  For small $\s$, i.e. light mass, and $k=0$ our bound decays with $\l$ like
  $\sim e^{\s\l}$, so that group averaging converges for any smooth states with $N >  -d/\s$ particles. This agrees with the threshold $N_{conv}$
  found in the previous section for finite-superposition states.

\end{itemize}
We will discuss the status of states with smaller numbers of particles
in section \ref{disc}.

In the remainder of this section we derive the bound on the finite boost
matrix elements $B_{j m k}(\l)$ discussed above. From the definition of
the boost matrix elements (\ref{eq:definitionOfFiniteMatrix}) we have
\eqn{ \label{eq:bound1}
  B_{j m k}(\l)
  &:=& \KG{ B(\lambda) \varphi_{\sigma \vj}}{ \varphi_{\sigma \vm} }
  \nn \\
  &=& -2 i \left[ \d_t \overline{\chY}_{\s m}(t) \right]
  \int_0^\pi d\chi (\sin \chi)^{d-1}\,
  \cYd_{m k}(\chi) \cYd_{j k}(\chi') \chY_{\s j}(t')
  \bigg|_{t = 0} .
}
In the second line we have integrated over the $S^{d-1}$ leaving only the
$t$- and $\chi$-dependant parts of the modes, and have also used the fact
that when evaluated on positive frequency states the two terms in the
Klein-Gordan inner product are equal. The function $\cYd_{j k}(\chi)$ is
defined in (\ref{eq:scalarSH}); the function $\chY_{\s j}(t)$ is defined in
(\ref{eq:chY}). To simplify the remaining integral over $\chi$ we will bound
both $\cYd_{j k}(\chi)$ and $\chY_{\s j}(t)$ in turn.

We begin by writing $\cYd_{j k}(\chi)$ in terms of Gegenbauer polynomials
$C_a^b(x)$:
\eqn{ \label{eq:cY}
  \cYd_{j k}(\chi)
  &=& \frac{\G(2k+d-1)}{2^{k+(d-1)/2}\G(k+\frac{d}{2})}
  \left[\frac{(2j+d-1)\G(j-k+1)}{\G(j+k+d-1)}
  \right]^{1/2}
  (\sin\chi)^k C_{j-k}^{k+(d-1)/2}(\cos\chi) . \nn \\
}
We may bound the size of $\cYd_{j k}(\chi)$ via the addition theorem of
Gegenbauer polynomials \cite{Vilenkin:1991}, which may be written
\eqn{ \label{eq:addn}
  C_j^p( \cos\theta \cos\chi &+& \sin \theta \sin \chi \cos\psi)\nn \\
  &=& \sum_{k=0}^j
  2^{2k} (2p+2k-1) \GGG{p+m,p+m,2p-1,1+j-m}{p,p,2p+j+k} \nn \\ & &
  \phantom{\sum_{k=0}^j\;}  \times
  (\sin \chi \sin \theta)^k C_{j-k}^{k+p}(\cos\chi)C_{j-k}^{k+p}(\cos\theta)
  C_k^{p-1/2}(\cos\psi) .
}
If we set $\psi = 0$ and $\theta = \chi$ then each term is positive
and we have that
\eq{ \label{eq:Gegbound}
  \left[(\sin \chi)^k C_{j-k}^{k+p}(\cos \chi)\right]^2
  \le \frac{1}{2^{2k}(2p+2k-1)}
  \GGG{p,p,2p+j+k}{p+m,p+m,2p-1,1+j-m} \frac{C_j^p(1)}{C^{p-1/2}_k(1)} ,
}
valid for $0 \le k \le j$, $p > 1/2$. From this expression we can derive
two useful bounds on $\cYd_{jk}(\chi)$. First, setting $k \to 0$ and
$p \to (d-1)/2$ in (\ref{eq:Gegbound}) we find
\eqn{ \label{eq:cYdbound0}
  |\cYd_{j,k=0}(\chi)| &\le&
  \left\{
    \frac{(2j+d-1)\G(j+d-1)}{2^{d-1}\G(1+j)\G\left(\frac{d}{2}\right)^2}
  \right\}^{1/2} .
}
This expression has technically been derived only for $d > 2$ (recall
we must have $p > 1/2$ in (\ref{eq:Gegbound})); however, (\ref{eq:cYdbound0})
turns out to be valid for $d=2$ as well. Second, for the case of
$k \neq 0$ it is useful to have a bound that includes some remnant
$\chi$-dependence. For this case we let
$p \to (d-1)/2 +1$, $k \to k-1$, $j \to j-1$ in (\ref{eq:Gegbound}) and
find
\eqn{ \label{eq:cYdboundk}
  |\cYd_{j,k>0}(\chi)| &\le&
  \left\{
    \frac{(2j+d-1)}{\sqrt{\pi}(2k+d-2) d }
    \GGG{k,j+d,\frac{d+1}{2}}{j,k+d-1,\frac{d}{2}}
  \right\}^{1/2} (\sin\chi) .
}
This expression is valid for $d \ge 2$ as well.

Next we examine the time-dependant part of the modes $\chY_{\s j}(t)$.
First, from (\ref{eq:chY}) we compute
\eq{ \label{eq:Aconst}
  \left[ \d_t \overline{\chY}_{\s m}(t) \right]_{t=0}
  = \frac{i \sqrt{\pi} N_{d \s m}}{2^{m+(d-3)/2} \G\left(\frac{m-\s}{2}\right)
    \G\left(\frac{m+\s+d}{2}\right)} .
}
To bound $\chY_{\s j}(t')$ we must bound the Legendre function
$P_{\s+(d-1)/2}^{-(j+(d-1)/2)}(i \sinh t)$. We may define the Legendre
function through the integral representation
\eq{ \label{eq:Legendre}
  P_\nu^\mu(z)
  = \frac{2^{-\nu} (z^2-1)^{\mu/2}}{\Gamma(-\mu-\nu)\G(\nu+1)}
  \int_0^\infty dx (z + \cosh x)^{\mu-\nu-1} (\sinh x)^{2\nu+1}
  \quad \Re (- \mu ) > \Re (\nu ) > -1 ,
}
for $\Re (- \mu ) > \Re (\nu ) > -1$; using this we bound
\eq{
  |\chY_{\s j}(t)| \le
  \frac{N_{d \s j}}{\left|2^{\s+(d-1)/2}\GG{\s +\frac{d+1}{2},j-\s}\right|}
  I
}
where
\eq{
  I := (\cosh t)^{-(\Re\,\s+d)}
  \int_0^\infty dx \left( 1 +\sech^2 t\sinh^2 x\right)^{-(j+\Re\,\s+d)/2}
  (\sinh x)^{2 \Re\,\s+d} .
}
For $\s > -d/2$ (corresponding to the complementary series) it is easy
to bound $I$ by 
$(\sinh x)^{2 \s+d} \le (\cosh x)(\sinh x)^{2\s+d-1}$ and using
the identity:
\eq{
  \int_0^\infty (\cosh x)(\sinh x)^{2a-1} (1+z \sinh^2 x)^{-a-b}
  = \half z^{-a} \GGG{a,b}{a+b}
}
for $\Re\, a > 0$, $\Re\,b > 0$, $z > 0$, yielding
\eqn{ \label{eq:chYbound1}
  |\chY_{\s j}(t) | &=& \frac{N_{d \s j}}{2^{\s+(d+1)/2}}
  \GGG{\frac{j-\s}{2},\s+\frac{d}{2}}{j-\s,\frac{j+\s+d}{2},\s + \frac{d+1}{2}}
 (\cosh t)^\s , \quad \left(\s > -\frac{d}{2}\right) .
}
For the case $\Re\,\s = -d/2$ (corresponding to the principal series)
we must bound $I$ differently; here
\eq{
  I = (\cosh t)^{-d/2}
  \int_0^\infty dx
  \left(1 + \sech^2 t \sinh^2 x\right)^{-\left(j+\frac{d}{2}\right)/2} .
}
Now note that
\eq{
  \left(1 + \sech^2 t \sinh^2 x\right) \le \half + \frac{1}{4}e^{2(x-t)} ,
}
so that
\eq{
  I \le (\cosh t)^{-d/2} \int_0^\infty dx
  \left(\half + \frac{1}{4}e^{2(x-t)}\right)^{-1/2}
  = \sqrt{2} (\cosh t)^{-j-d/2} {\rm arcsh}(\sqrt{2} e^t) ,
}
and we have the bound
\eqn{ \label{eq:chYbound2}
  |\chY_{\s j}(t)| &\le&
  \frac{ N_{d \s j}}{| 2^{\s+d/2-1} \GG{j-\s,\s+\frac{d+1}{2}} |}
  (\cosh t)^{-d/2} {\rm arcsh}(\sqrt{2} e^t) ,
  \quad \left( \Re\, \s = -d/2 \right) . \nn \\
}
We conclude by combining (\ref{eq:chYbound1}) and (\ref{eq:chYbound2});
because we are primarily interested in the behavior of $B_{j m k}(\l)$ at
large $j \gg 1$, we take the liberty of assuming $(j - \Re\,\s )\ge 3$
which allows us to simplify the expression slightly:
\eqn{ \label{eq:chYbound}
  |\chY_{\s j}(t)|
%   &\le& \frac{\sqrt{\pi} N_{d \s j} }{2^{j+(d-3)/2}
%     \G\left(\frac{j-\s}{2}\right)\G\left(\frac{j+\s+d}{2}\right)
%   \left|\G\left(\s+\frac{d+1}{2}\right)\right|} \nn \\ & & \times
%   (\cosh t)^{\Re \,\s}
%    \left\{
%      \begin{array}{c}
%        \half \G\left(\s+\frac{d}{2}\right) \\
%        \sqrt{2}\, {\rm arcsh}( \sqrt{2}\, e^t )
%      \end{array} \right\} \quad
%      \begin{array}{l}
%        {\rm for \; } \s > -d/2 \\
%        {\rm for \; } \s = -d/2+i \rho
%      \end{array}  .
  &\le& \frac{1}{|\G\left(\s+\frac{d+1}{2}\right)|}
  \GGG{\frac{j-\s+1}{2},\frac{j+\s+d+1}{2}}
  {\frac{j-\s}{2},\frac{j+\s+d}{2}}^{1/2}
  (\cosh t)^{\Re \,\s} \nn \\ & & \times
   \Bigg\{
     \begin{array}{c}
       \half \G\left(\s+\frac{d}{2}\right) \\
       \sqrt{2}\, {\rm arcsh}( \sqrt{2}\, e^t )
     \end{array}  \quad
     \begin{array}{l}
       {\rm for \; } \s > -d/2 \\
       {\rm for \; } \s = -d/2+i \rho
     \end{array}  .
}
Note that in this expression the $t$- and $j$- dependence separate.

We now insert our bounds (\ref{eq:cYdbound0}), (\ref{eq:cYdboundk}),
and (\ref{eq:chYbound})
into our expression for the finite boost matrix elements (\ref{eq:bound1}).
From the coordinate transformation (\ref{eq:coordinateChange}) we see that
at $t = 0$ we have $(\cosh t')^2 = 1 +\sinh^2\l \cos^2\chi$ and that
$t' \le \l$; using these facts one may evaluate the integral (\ref{eq:bound1}),
finding the bounds
\eq{
  | B_{j,m,k=0}(\l)| \le f(j, m, k=0) g_0(\l), \quad
  | B_{j,m,k>0}(\l)| \le f(j, m, k>0) g_1(\l), \quad
}
where
\eqn{
  f(j,m,k=0) &:=& \frac{1}{|\G\left(\s +\frac{d+1}{2}\right)|}
  \left\{
   \GGG{\frac{j-\s+1}{2},\frac{j+\s+d+1}{2},\frac{m-\s+1}{2},\frac{m+\s+d+1}{2}}
   {\frac{j-\s}{2},\frac{j+\s+d}{2},\frac{m-\s}{2},\frac{m+\s+d}{2}}
  \right\}^{1/2} \nn \\ & & \times
  \frac{\sqrt{\left(j+\frac{d-1}{2}\right)\left(m+\frac{d-1}{2}\right)}}
  {\G\left(\frac{d}{2}\right)}
  \left\{
    \GGG{j+d-1,m+d-1}{1+j,1+m}
  \right\}^{1/2} , \\
  f(j,m,k>0) &:=&
  \frac{ 1 }{|\G\left(\s +\frac{d+1}{2}\right)|}
  \left\{
   \GGG{\frac{j-\s+1}{2},\frac{j+\s+d+1}{2},\frac{m-\s+1}{2},\frac{m+\s+d+1}{2}}
   {\frac{j-\s}{2},\frac{j+\s+d}{2},\frac{m-\s}{2},\frac{m+\s+d}{2}}
  \right\}^{1/2} \nn \\ & & \times
  \frac{\G(k)\sqrt{\left(j+\frac{d-1}{2}\right)\left(m+\frac{d-1}{2}\right)}}
  {(d+1)(2k+d-2)\G(k+d-1)}
  \left\{
    \GGG{j+d,m+d}{j,m}
  \right\}^{1/2} ,
}
\eqn{
  g_0(\l) &:=&
  (\cosh \l)^{\Re\,\s}
  \2F1{-\frac{\Re\,\s}{2}}{\frac{d}{2}}{\frac{d+1}{2}}{\tanh^2\l} \nn \\
  & & \times
  \Bigg\{
    \begin{array}{c}
      \half \G\left(\s+\frac{d}{2}\right) \\
      \sqrt{2}\, {\rm arcsh}( \sqrt{2}\, e^{\l} )
    \end{array} \quad
    \begin{array}{l}
      {\rm for \; } \s > -d/2 \\
      {\rm for \; } \s = -d/2+i \rho
    \end{array}, \\
  g_1(\l) &:=&
  (\cosh \l)^{\Re\,\s-1}
  \2F1{\frac{1-\Re\,\s}{2}}{\frac{d+2}{2}}{\frac{d+3}{2}}{\tanh^2\l} \nn \\
  & & \times
  \Bigg\{
    \begin{array}{c}
      \half \G\left(\s+\frac{d}{2}\right) \\
      \sqrt{2}\, {\rm arcsh}( \sqrt{2}\, e^{\l} )
    \end{array} \quad
    \begin{array}{l}
      {\rm for \; } \s > -d/2 \\
      {\rm for \; } \s = -d/2+i \rho
    \end{array} .
}
Applying Stirling's approximation to the gamma functions confirms that $f(j,m,k)$ is bounded by a polynomial in $j$ as desired. The asymptotics
of $g(\l)$ state in (\ref{g}) follow from the well-known asymptotics of
the hypergeometric function \cite{GR,Bateman,Slater}.

%%%%%%%%%%%%%%%%%%%%%%%%%%%%%%%%%%%%%%%%
\section{Higher Spin}\label{HS}
%%%%%%%%%%%%%%%%%%%%%%%%%%%%%%%%%%%%%%%%

The techniques of section \ref{sec:scalar} can also be used to study
the convergence of group averaging for linear vector and tensor gauge
fields. In fact, with the appropriate choice of gauge and in the
appropriate basis, the matrix elements of the infinitesimal boost generator $\hat B$ will
turn out to have precisely the same form as in the scalar case.  As a result,
we can use our results from section \ref{sec:scalar} to evaluate the matrix
elements of $B(\lambda)$ between vector and tensor one-particle states
and to quickly arrive at the desired result.

\subsection{Massless Vector Fields}
\label{sec:vector}

Massless vector fields obey the Maxwell equation of motion
\eq{ \label{eq:vectorEOM}
  \Box A_\mu - \frac{d}{\ell^2} A_\mu -  \nabla_\mu \nabla^\nu A_\nu = 0\;,
}
which is of course invariant under the gauge transformation
\eq{ \label{eq:vectorGaugeFreedom1}
  A_\mu \to A_\mu + \ell \nabla_\mu \varphi \;,
}
for any scalar $\varphi$.  Note that
$ \nabla^\mu A_\mu \to \nabla^\mu A_\mu + \ell \Box \varphi$, so
choosing $\varphi = \ell^{-1} \Box^{-1} \nabla^\mu A_\mu$ brings us to a gauge in
which  $A_\mu$ satisfies
\eq{ \label{eq:masslessVectorEOM}
  \left(\Box - \frac{d}{\ell^2}\right) A_\mu = 0, \quad
  \nabla^\mu A_\mu = 0\;.
}
There remains some residual gauge symmetry which we will discuss below.
As in the scalar case it is convenient to define a quantity
$\sigma_1$ related to the eigenvalue of the wave operator.
For vector fields the natural choice \cite{Higuchi:1986wu}
is
\eq{ \label{eq:definitionOfSigma1}
  \Box A_\mu = \left[ \frac{-\sigma_1(\sigma_1+d)+1}{\ell^2}\right] A_\mu\; ,
}
so that $\sigma_1 = -1$ for the massless case.

Transverse vector fields $A_\mu$ satisfying (\ref{eq:definitionOfSigma1}) in
$dS_D$ are the analytic continuation of
transverse vector spherical harmonics on $S^D$ (reviewed in
Appendix~\ref{app:vector}). There are $D-1$ sets of independent modes,
corresponding to the fact that each set of modes has non-vanishing
components only for indices tangent to the sub-sphere $S^2, S^3,\dots,S^D$.
We label the sets of modes $A_\mu^{(2;\vec j)}, \dots, A_\mu^{(d;\vec j)}, A_\mu^{(D;\vec j)}$, where $\vec j$ is as in \eqref{eq:js}.

As mentioned above, there is some residual gauge freedom in
(\ref{eq:masslessVectorEOM}); namely, there remains the freedom
$A_\mu \to A_\mu + \ell \nabla_\mu \varphi_{0}$ for $\varphi_0$ satisfying
the Klein-Gordon equation with $\sigma = 0$. In appendix
\ref{residual} we use this freedom to eliminate the modes
$A_\mu^{(D;\vj)}$. Since $A_t$ vanishes for all other modes, we have fixed temporal gauge
$A_t = 0$. The remaining modes are
\eqn{
  \label{eq:Aba}
  A_a^{(\beta;\vj)} &=& \ell^{(3-d)/2}
  (\cosh t) \,\chY_{\sigma_1 j}(t) \, \Vd_a^{(\beta;\vj)}(\Omega_d) .
}
Here $a$ is a spatial index, $\beta = 2,\dots,d$ denotes the set of modes,
and $\Vd_a^{(\beta;\vj)}(\Omega_d)$ are vector harmonics on $S^d$
(see Appendix~\ref{app:vector}). Note that such harmonics exist only for $j \ge k \ge 1$.

As with our scalar modes, these modes can be shown to satisfy a de
Sitter invariant positive frequency condition.  In addition, they have
been normalized with respect to the inner product
\eq{
  \KG{A^{(\alpha;\vj)}}{A^{(\beta;\vm)}}
  = -i \int d\Sigma^\mu A^{(\alpha;\vj)}_\nu \overleftrightarrow{\nabla}_\mu
  \overline{A}^{(\beta;\vm)\,\nu}
  = \delta^{\alpha\beta}\delta_{\vj \vm}\;,
  \label{eq:vectorInnerProduct}
}
which agrees with the symplectic structure of the Maxwell field
when $A_t=0$ as above.

We therefore expand the field in creation and annihilation operators
$a^{\dagger(\beta;\vj)}$ and $a^{(\beta;\vj)}$ as
\eq{
  A_\mu(x) = \sum_{\beta}\sum_{\vj}
  \left[ a^{(\beta;\vj)} A^{(\beta;\vj)}_\mu(x)
    + a^{\dagger(\beta;\vj)} \overline{A}^{(\beta;\vj)}_\mu(x)
  \right]\;.
}
It follows from  (\ref{eq:vectorInnerProduct}) that the creation and annihilation operators
have the usual commutators
\eq{
  \com{a^{(\alpha;\vj)}}{a^{\dagger(\beta;\vm)}}
  = \delta^{\alpha\beta}\delta_{\vj\vm}, \quad
  \com{a^{(\alpha;\vj)}}{a^{(\beta;\vm)}}
  = \com{a^{\dagger(\alpha;\vj)}}{a^{\dagger(\beta;\vm)}}
  = 0\; ,
}
and allow us to define an auxiliary space vacuum state which satisfies
 \eq{
  a^{(\beta;\vj)} \dvac = 0 \quad \forall\; \beta,\;\vj\;.
}
General states in $\cHa$ are linear combinations of the states
$a^{(\beta;\vj_1)}\cdots a^{(\beta_n;\vj_n)}\dvac$.

%%%%%%%%%%%%%%%%%%%%%%%%%%%%%%%%%%%%%%%%
\subsection{Vector Boost Matrix Elements}
%%%%%%%%%%%%%%%%%%%%%%%%%%%%%%%%%%%%%%%%

We now compute the matrix elements of the boost generator $\hB$ that
generates boosts along $\hB^\mu$ (\ref{eq:boostVector}). The action
of $\hat B$ is given by
\eq{
  \com{\hB}{A_\mu} = i \Lie_B A_\mu - \PG\;,
}
where $\PG$ denotes the subtraction of a ``pure gauge'' term required
to maintain the above gauge conditions. In particular, this subtraction
is required to maintain the temporal gauge condition $A_t=0$ which breaks
manifest de Sitter invariance. It is convenient denote the action of a
boost on the modes by $\delta_B$ so that
\eq{
  \com{\hB}{A_\mu} = \sum_\beta \sum_{\vj}
  a^{(\beta;\vj)} \delta_B A_\mu^{(\beta;\vj)} + {\rm h.c.}
}
In general, this action takes the form
\eq{ \label{dbeq}
  \delta_B A_\mu^{(\beta;\vj)}
  =  i \Lie_{B} A_\mu^{(\beta;\vj)} - c \ell \nabla_\mu \varphi_{0,\vj} \;,
}
where $c$ is a constant.

Let us first consider the modes $A_\mu^{(\beta;\vj)}$ for $\beta \neq d$.
Since these modes vanish in the subspace of the co-tangent space
associated with $B^\mu$
(i.e., $A_t^{(\beta;\vj)} =0$ and $A_\chi^{(\beta;\vj)} = 0$), we also have
\eq{
  \Lie_{B} A_t^{(\beta;\vj)} =  \Lie_{B} A_\chi^{(\beta;\vj)} = 0\;.
}
Thus, the Lie derivative preserves temporal gauge when acting on such
modes and there  is no pure gauge contribution; the action of $\delta_B$
is given simply by $ i \Lie_B$.  After a short calculation one finds
\eq{
  \delta_B A_\mu^{(\beta;\vj)} =
  b^+_{-1,\vj,} A_\mu^{(\beta;\vjp)}
  + b^-_{-1,\vj,} A_\mu^{(\beta;\vjm)}\;,
}
where the coefficients are identical to those for scalars (\ref{eq:bpm})
with $\sigma =-1$. These modes ``transform like scalars'' in the sense
that they have boost matrix elements identical to scalars
with $\sigma = -1$ (as one might expect for massless vector fields).

It remains only to consider the modes $A_\mu^{(d;\vj)}$. Once again, the
details are presented in appendix \ref{residual} where it is shown that
$\delta_B  A_\mu^{(d;\vj)}$ takes the form (\ref{dbeq}) with
\eq{
   c \ell \nabla_\mu \varphi_{0,\vj}
  = \frac{i}{j(j+d-1)} \nabla_\mu
  \left[\left(\cosh t \,\d_t + (d-1)\sinh t\right)
    \frac{\sin\chi}{\cosh t} A_\chi^{(d;\vj)} \right]\;.
}
It is now straightforward to compute matrix elements of $\hB$ for the
$A_\mu^{(\chi;\vj)}$ modes.  A key point is that the Lie algebra requires  $\delta_B A_\mu^{(d;\vj)}$ to be a linear combination of
$A_\mu^{(d;\vjp)}$ and  $A_\mu^{(d;\vjm)}$, so that we need only determine
the coefficients.  This is most easily done by considering the single
component $A_\chi^{(d;\vj)}$.   After some tedious algebra one finds
\eq{
  \delta_B A_\chi^{(d;\vj)} = b^+_{-2,j}A_\chi^{(d;\vjp)}
  + b^-_{-2,\vj}A_\chi^{(d;\vjm)}\; ,
}
where the coefficients are identical to those for scalars (\ref{eq:bpm})
with $\sigma =-2$.

For each set of modes above, we found that $\hat B$ has precisely the same
matrix elements as for scalar fields with either $\sigma =-1$ or
$\sigma =-2$. As a result, the matrix elements of finite boosts
$B(\lambda)$ can be read directly from
\footnote{Note that for $D=4$ (i.e., $d=3$) the matrix elements with
$\sigma=-1$ and $\sigma=-2$ are identical as a result of the invariance
under $\sigma \to - (\sigma +d)$, and are identical to the matrix
elements of a conformally coupled scalar field. They therefore
reduce to the simplified expression (\ref{eq:conformal}).} (\ref{eq:solution}) and the
group-averaging properties follow immediately. The results for
vector fields are thus:

\begin{itemize}

\item{} ${\bf A_\mu}$$^{(d;\vj)}$: These modes act like scalars with
$\sigma=-2$. Thus the group averaging expression $\cI$ (\ref{noproj})
for $N$-particle states built from a finite number of such wavefunctions
converges when $d - 2 N - \sum_i^N k_i < 0$. Since the minimum value
of $k$ for these modes is $k=1$, this generally requires $N > N_{conv} = d/3$.

\item{} ${\bf A_\mu}$$^{(\beta;\vj)}{\bf\;for\;\beta\neq d}$:
These modes act like scalars with $\sigma=-1$. Thus, for $N$-particle states
built from a finite number of such wavefunctions, $\cI$ converges when
$d -  N - \sum_i^N k_i < 0$. Since the minimum value of $k$ is $k=1$,
convergence generally requires $N > N_{conv} = d/2$.

\item{} {\bf Both:} For $N$-particle states built from a finite number of  wavefunctions of either type,  $\cI$ again converges when
$N > N_{conv} = d/2$.  It is also clear that $\cI$ may converge for
$N < N_{conv}$ if the particles have high enough angular momentum $k$ and/or mostly have $\beta = d$.

\item{} {\bf Arbitrary smooth states:} An arbitrary smooth wavefunction
may be written as an infinite superposition of the above modes, each of which has
$\s \le -1$. From our discussion in section \ref{sec:bound} we conclude
that group averaging converges for general smooth states when $N >  d$.
\end{itemize}

%%%%%%%%%%%%%%%%%%%%%%%%%%%%%%%%%%%%%%%%
\subsection{Gravitons} \label{sec:graviton}
%%%%%%%%%%%%%%%%%%%%%%%%%%%%%%%%%%%%%%%%

We now address group averaging for free gravitons, following the same
recipe as in the preceding sections\footnote{Group averaging of gravitons for $d=3$ was considered in
\cite{AHII}.  For this special case, one can obtain more detailed
results than we derive below.}.
After quantizing metric perturbations to create the auxiliary Hilbert
space $\cHa$, we compute the infinitesimal boost generators for
gravitons and find that they again agree with expressions found in
the scalar case.  Thus, as in the case of vector fields, the
results of section \ref{sec:scalar} allow us to quickly arrive at conditions
under which group averaging will converge.

We wish to consider linearized perturbations $h_{\mu\nu}$ of the
background de Sitter metric $g_{\mu\nu}$.
Starting from the Einstein-Hilbert action
$ S = \int d^D x \sqrt{-g}(R - 2 \Lambda)$,
one finds the linearized equation of motion
\eqn{ \label{eq:gravitonEOM1}
  0 &=&  \Box h_{\mu\nu} + \nabla_{\mu} \nabla_{\nu} h
  - \nabla_\mu \nabla^\lambda h_{\lambda \nu}
  - \nabla_\nu \nabla^\lambda h_{\lambda \mu}
  - \frac{2}{\ell^2} h_{\mu\nu} \nn \\ & &
  + \left( \nabla^\alpha\nabla^\beta h_{\alpha\beta}
    - \Box h
    + \frac{3-D}{\ell^2} h \right) g_{\mu \nu} \;,
}
where $h = g^{\mu\nu}h_{\mu\nu}$ and once again
$\Box =  \nabla^\mu\nabla_\mu$ is the wave operator on de Sitter.
We impose transverse traceless gauge
\eq{
  \nabla_\lambda h^{\lambda \mu}  =0 , \ \ \ h = 0,
}
so that the equations of motion become
\eq{ \label{eq:gravitonEOM3}
  \left(\Box - \frac{2}{\ell^2} \right) h_{\mu\nu} = 0\;, \quad
  \nabla^\lambda h_{\lambda\mu} = h = 0\;.
}

Not surprisingly, mode solutions to (\ref{eq:gravitonEOM3}) are
transverse traceless tensor spherical harmonics on $S^D$ analytically
continued to $dS_D$. Transverse traceless tensor harmonics are reviewed
in Appendix~\ref{app:tensor}. After analytic continuation, the general such
harmonics satisfy
\eq{
  \Box h_{\mu\nu}
  = \left[\frac{-\sigma_2(\sigma_2+d)+2}{\ell^2} \right] h_{\mu\nu}\;,
}
in addition to the transverse-traceless condition. Solutions
to our equation of motion (\ref{eq:gravitonEOM3}) are those with
$\sigma_2 =0$. There are $(D+1)(D-2)/2$ sets of independent modes
with non-vanishing components tangent to the sub-spheres $S^2,\dots,S^D$.
We label the sets of modes $h^{(\alpha\beta)}$ where
$\alpha,\beta = 2,3,\dots,d,D$.
Individual modes are then labeled by momenta $\vj$.

Recall that for free vector fields imposing transverse gauge still allowed the residual gauge symmetry
$A_\mu \to A_\mu + c \ell \nabla_\mu \varphi_{0}$,  and that used
this symmetry to set $A_t = 0$ (temporal
gauge). A similar story holds for gravitons. The equations of
motion (\ref{eq:gravitonEOM3}) are invariant under the gauge
transformation $h_{\mu\nu} \to h_{\mu\nu} + \ell \nabla_{(\mu} A_{\nu)}$
so long as $A_\mu$ obeys
\eq{ \label{eq:AGaugeFreedom}
  \left(\Box + \frac{d}{\ell^2} \right) A_\mu = 0\;, \quad
  \nabla^\mu A_\mu = 0\;.
}
Following the same steps as in the vector case
(see Appendix \ref{residual}) one can show that the tensor
modes $h^{(D\alpha)}_{ab}$, $\alpha = 2,\dots,D$ are in fact
proportional to $\nabla_{(\mu} A_{\nu)}$, and can therefore
be set to zero, fixing the gauge completely. Since these are the only modes with non-vanishing $t$ component, this amounts
to fixing $h_{t\mu} = 0$.

The $D(D-3)/2$ remaining sets of modes are
\eqn{
  h^{(\alpha\beta;\vj)}_{t\mu} &=& 0\;, \nn\\
  h^{(\alpha\beta;\vj)}_{ab} &=& \ell^{(5-d)/2}
  (\cosh t)^2 \, \chY_{0,j}(t) \, \Td_{ab}^{(\alpha\beta;\vj)}(\Omega_d),
  \quad \alpha,\;\beta = 2,\dots,d\; ,
}
where $\Td_{ab}^{(\alpha\beta;\vj)}(\Omega_d)$ are tensor harmonics on
$S^d$ (see Appendix~\ref{app:tensor}). Note that rank two tensor
harmonics exist only for
$j_d \ge j_{d-1} \ge \dots \ge j_2 \ge |j_1| \ge 2$.
As in the scalar and vector cases considered before, these modes satisfy
a de Sitter invariant positive frequency condition. The modes have been
normalized with respect to the inner product
\eq{ \label{eq:gravitonInnerProduct}
  \KG{h^{(1)}}{h^{(2)}}
  :=  -i \int d\Sigma^\lambda\,
  h^{(1)}_{\mu\nu} \,\overleftrightarrow{\nabla_\lambda}
  \,\overline{h^{(2)}}^{\mu\nu}\;.
}

We can therefore expand $h_{\mu\nu}$ in creation and annihilation
operators
\eq{
  h_{\mu\nu}(x) = \sum_{\alpha}\sum_{\beta}\sum_{\vj}
  \left[ a^{(\alpha\beta;\vj)} h_{\mu\nu}^{(\alpha\beta;\vj)}(x) +
    a^{\dagger(\alpha\beta;\vj)} \overline{h}_{\mu\nu}^{(\alpha\beta;\vj)}(x)
  \right]
}
and complete the canonical quantization of $h_{\mu\nu}$ much as was done above for $\varphi$ and $A_\mu$; see  \cite{Antoniadis:1986sb, Allen:1986tt,Higuchi:1991tk,Higuchi:1991tn,
Higuchi:2001uv} for useful references.

%%%%%%%%%%%%%%%%%%%%%%%%%%%%%%%%%%%%%%%%%%%%%%%%%%%%%%%%%%%%%%%%%
\subsection{Graviton boost matrix elements}
%%%%%%%%%%%%%%%%%%%%%%%%%%%%%%%%%%%%%%%%%%%%%%%%%%%%%%%%%%%%%%%%%

Finally, we compute the finite boost matrix elements
for gravitons. The calculation is essentially the
same as for free vector fields, so we will simply summarize
the results.

The boost generator $\hB$ acts on the graviton field as
\eq{
  \com{\hB}{h_{\mu\nu}} = i \Lie_B h_{\mu\nu} - \PG \;,
}
where once again $\PG$ denotes a pure gauge term that must be
subtracted in order to preserve temporal gauge $h_{t\mu}=0$.
The pure gauge term must be proportional to $\ell \nabla_{(\mu} A_{\nu)}$
where $A_\mu$ satisfies (\ref{eq:AGaugeFreedom}).
The calculation of this pure gauge term is analogous
to the calculation done for free vector fields presented
in appendix \ref{residual}. One finds that all modes transform
like
\eq{
  \delta_B h^{(\alpha\beta;\vj)}_{\mu\nu}
  = b^+_{\sigma\vj} \, h^{(\alpha\beta;\vjp)}_{\mu\nu}
  + b^-_{\sigma\vj} \, h^{(\alpha\beta;\vjm)}_{\mu\nu}
}
with the following values of $\sigma$:

\begin{itemize}

\item ${\bf h_{\mu\nu}}^{(d d;\vj)}$: these modes
  transform with a value of $\sigma = -2$. Thus, for $N$-particle states built from a finite number of such wavefunctions , $\cI$ (\ref{noproj})
  converges when $d - 2 N - \sum_i^N k_i < 0$. Since the minimum value
  of $k$ for tensor modes is $k = 2$, this requires
  $N > N_{conv} =  d/4$. These are the most convergent modes.

\item ${\bf h_{\mu\nu}}^{(d \alpha;\vj)},\;{\bf \alpha\neq d}$: these modes
  transform with a value of $\sigma = -1$. Thus, for $N$-particle states built from a finite number of such wavefunctions, $\cI$ converges
  for $N > N_{conv} =  d/3$.

\item ${\bf h_{\mu\nu}}^{(\alpha\beta;\vj)}{\bf,\;\alpha,\beta \neq d}$:
  these modes transform with a value of $\sigma = 0$. Thus, for $N$-particle states built from a finite number of such wavefunctions, $\cI$
  converges for $N > N_{conv} =  N_{conv} = d/2$. These are the least convergent
  modes.

\item {\bf Arbitrary smooth states}: For all modes the quantity
$\s - k <  -1$, so following the discussion in section \ref{sec:bound}
we conclude that group averaging will converge for arbitrary smooth states when
$N >  d$.

\end{itemize}

As in the vector case, modes orthogonal to $B^\mu$ are the least
convergent while those with non-zero overlap are more convergent.
For finite superposition states, the number of particles required
for convergence is set by the least convergent modes to be
$N > N_{conv}= d/2$. It is again useful to note that for $D=4$
(i.e., $d=3$) the finite boost matrix elements are the same for
$\sigma=-2$ and $\sigma=-1$ and are given by the simplified
expression (\ref{eq:conformal}).

%%%%%%%%%%%%%%%%%%%%%%%%%%%%%%%%%%%%%%%%%%%%%%%%%%%%%%%%%%%%%%%%%
\section{Discussion} \label{disc}
%%%%%%%%%%%%%%%%%%%%%%%%%%%%%%%%%%%%%%%%%%%%%%%%%%%%%%%%%%%%%%%%%

We have studied the general behavior of matrix elements in a standard basis
for free massive scalars, abelian vector gauge fields, and linearized
gravitons in any dimension $d$. We have shown that group-averaging
converges for states with $N$-particle wavefunctions given by de Sitter 
harmonics -- as well as finite superpositions of such basis states -- when 
all components of the quantum state contain $N > N_{conv}$ particles. For 
principal series scalars $N_{conv} = 2$ (so we require $N \ge 3$), for 
scalars in the complementary series and in the discrete series with 
$M^2 \neq 0$ we have $N_{conv} = d/|\s|$, while for both vector gauge 
fields and linearized gravitons $N_{conv}= d/2$.

We also considered states with general smooth wavefunctions. Such states 
are in general infinite superpositions of the above basis states. Due to 
the difficulty in computing the group averaging inner product exactly, our 
results for this case relied on finding a bound. The key step here to 
bounding the finite boost matrix elements $B_{j m k}(\l)$ as discussed in 
section \ref{sec:bound}. Our bound grows like a polynomial
in total angular momenta but decays with rapidity at a weaker rate than
the actual matrix elements. As a result, for principal-series scalars, linearized gauge fields, and linearized gravitons we can only demonstrate convergence for general smooth wavefunctions with $N > d$ particles, a
somewhat larger number than for finite superpositions of our de Sitter harmonics.
We have by no means shown that such a high particle content is necessary.
In fact, as discussed below, we suspect that the values of $N_{conv}$ stated
 above for finite superpositions should also suffice for states with
arbitrary smooth wavefunctions.

Our bound on the finite boost matrix elements grows with
angular momenta like a polynomial of order $\mathcal{O}((j m)^{(d+1)/2})$,
suggesting that the group averaging inner product should likewise grow
with angular momenta. However, in cases where the inner product
has been computed exactly (or numerically) it has been found that the inner
product actually \emph{decreases} with angular momenta. For example, Higuchi
\cite{AHII,AHnotes} computed the group averaging inner product exactly for massless $1+1$ scalars\footnote{See \cite{Marolf:2008} for a treatment of the zero-mode.},
conformally-invariant scalar fields in $3+1$ dimensions, and $3+1$ massless
tensor fields. All of these cases have $\s = - (d-1)/2$, i.e. the same as 
for a conformally coupled scalar field. For this special value of $\s$ it is 
possible to define a set of states in $\cHa$ which map under group averaging 
to a complete orthonormal basis in $\cHp$ for states with 
$N > d / \s = 2 d /(d-1)$ (and in the case of $d=1$ Higuchi finds $N > 1$); 
see also the discussion in \cite{Marolf:2008}. Higuchi's results show that 
the group averaging norm decreases with total angular momentum, so that 
group averaging in fact converges for any normalizable state in $\cHa$ with 
$N > 2 d /(d-1)$ (or $N>1$ for $d=1$).

{
\FIGURE{\includegraphics[width=5cm] {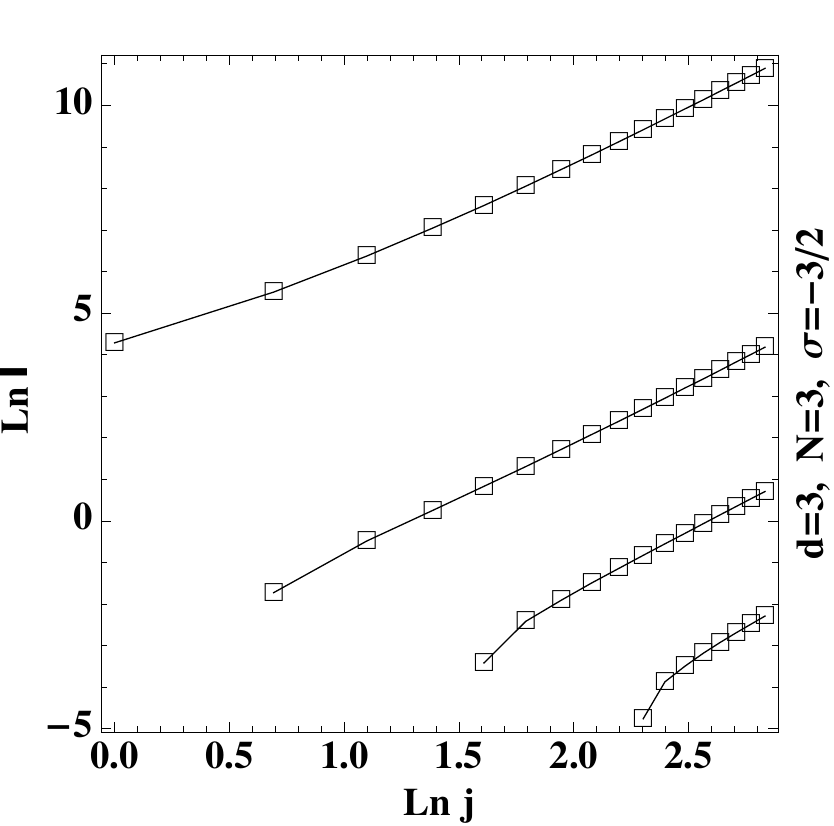} \ \ \ \ \ \ \ \ \includegraphics[width=4.75cm] {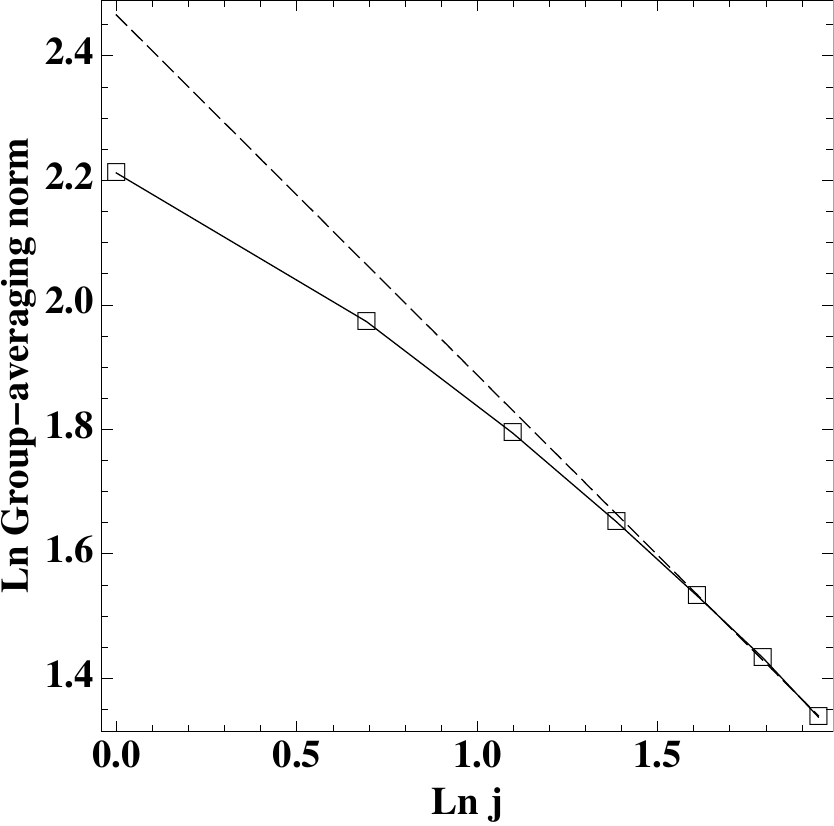}
  \caption{{\bf Left:} The group-averaging expression
    $\cI = \int d\l (\sinh \l)^d (B_{jjk}(\l))^N$. Shown (from top to bottom) are the cases
    $k=0,2,5,10$. {\bf Right:} The group-averaging norms of rotationally-invariant
    states. The dashed line is the best-fit
    curve $\sim j^{-1/2}$.  For both figures
    $N=3$, $d=2$, $\sigma = -3/2$
}
}

}

The difference between Higuchi's results and our own might
be attributed to i) the possibility that the solvable cases
$\s = - (d-1)/2$ may be exceptional, ii) some weakness of our bound on
the matrix elements $B_{j m k}(\l)$, or iii) the fact that we have not
considered the effect of the $SO(D)$ projection operators $\Proj$.
Numerical results suggest that (iii) is the most important effect.
Consider for example
a free $2+1$ scalar field with $\s = -3/2$ (at the border of the
principal and complementary series).
The quantity $\cI$ (the group averaging inner product
\emph{without} the $SO(3)$ projection operators) from
(\ref{noproj}) is shown in figure 1 (left) for a particular sequence of normalized states.
It is clear that  $\cI$ increases with $j$ (though notably not
as rapidly as our bound).  In contrast, one can also 
consider $\cI$ for a family of rotationally-invariant $3$-particle states of the form $| 2\ {\rm singlet}, j \rangle \otimes |1\ {\rm singlet}
\rangle$, where $|2\ {\rm singlet}, j \rangle$ is the unique
normalized rotationally invariant 2-particle state in which each
particle has total angular momentum $j$ and $|1\ {\rm singlet}
\rangle$ is the normalized rotationally-invariant one-particle
state. Because such states are already rotationally invariant, $\cI$ coincides with their group averaging norm.  As shown in figure 1 (right), these norms {\em decrease} with $j$,  and in particular appear consistent with a power law of the form $\sim j^{-1/2}$.  From this we conclude that
including the $SO(3)$ projection matrices is essential to obtaining the correct angular momentum dependence of the group averaging norm.
If indeed the group averaging inner product of normalized rotationally-invariant states decreases with $j$, then the above-stated values of $N_{conv}$ will
guarantee that group averaging converges for general normalizeable states.

We conclude by discussing the status of group averaging for
states with $N \le N_{conv}$.
Let us begin with the case  $N=0$. In all cases addressed above, the
field admits a de Sitter-invariant vacuum.  It is clear that the group
averaging norm diverges for this state, since
\eqref{eq:definitionOfPhysicalStates} would integrate
$\langle 0| U(g) | 0 \rangle = 1$ over the entire non-compact de Sitter group.  On the other hand, since $|0 \rangle$ is already de Sitter invariant,
there is no need to group average it to define a physical state.  This case
is directly analogous to examples studied in \cite{single} and the same
conclusion holds: If ${\cal O}$ is de Sitter invariant, then so is
${\cal O} |0 \rangle$.  But the only normalizable such state is the vacuum
itself.  Thus, under the action of de Sitter-invariant observables, one
finds that the state $|0 \rangle$ is superselected from all other states.
Since group averaging is a technique for taking the action of this
observable algebra on $\cHa$ and constructing a representation on de
Sitter-invariant states, it is clear that the vacuum must be treated
separately.  See \cite{single, ALMMT,Gomberoff:1998ms,JLAM03,JLSum,JLAM05}
for more discussion and further examples of this phenomenon.

Another interesting point is that group averaging never converges for 1-particle states.  One may see this from the results above by noting that all $SO(D)$-invariant 1-particle states have $j=0$ and also noting that  neither vector gauge particles nor linearized gravitons exist for $D=2$.    This lack of convergence for $N=1$ is a natural consequence of the linearization-stability constraints, as they suggest that no physical 1-particle states should exist.  In particular, at the classical level such constraints forbid the existence of 1-particle spacetimes.  By this we mean spacetimes which differ from de Sitter space only by a localized particle-like excitation carrying positive energy in the sense of the (timelike) weak energy condition.  The constraints state that the de Sitter charges $Q[\xi] = \int_\Sigma n^a \xi^b T_{ab}$ must vanish, while for a localized distribution of positive energy one may always choose $\xi^b$ to be timelike and future-pointing in the region where $T_{ab}$ is non-vanishing.  There is nothing to `balance' this charge on the opposite side of the $\xi^b-$Killing horizon (where $\xi^b$ becomes past-pointing or spacelike), so the constraints cannot be satisfied.

Though group-averaging of our 2-particle basis sates converges for the physically interesting cases of $D=4$ gravitons and vector fields, this is not generally true in higher dimensions or for massive scalars.  Now, classical 2-particle spacetimes do exist, but they are highly constrained as the two particles must `balance' exactly.  In fact, in the test-particle limit the worldline of either particle must be mapped to the other under the de Sitter antipodal map.  As a result, these worldlines preserve an $\mathbb{R} \times$ SO($D-1$) subgroup of the de Sitter group SO($D,1$).  The $\mathbb{R}$ factor is associated with a non-compact boost generator.      Thus, at least in the classical limit, the behavior of two-particle states is much like that of the vacuum in that they become invariant under a non-compact symmetry group.  This suggests that 2-particle states may also form a superselection sector under the observable algebra, and that some renormalized version of group averaging (see \cite{ALMMT,Gomberoff:1998ms}) might converge for such states.  While
dispersion of the wavefunction breaks the above symmetry in a quantum 2-particle state, it may be that this effect is generally small enough for the above analogy to hold.

Cases where group averaging fails for $N$-particle states with $N > 2$ may be similar.   We note that this occurs only for scalars in the complementary and discrete series and for our massless higher spin fields.  None of these cases have a  classical 1-particle limit since their Compton wavelength is the de Sitter scale or larger.

Though group averaging does not generally converge for $N \le N_{conv}$, one may use the $k$-dependence of \eqref{scalarConv} to show that group averaging {\it does} converge for certain carefully constructed such states.  The trick is to build a state for which no particle has $k=0$.  We do not attempt to classify such states, but merely present a simple example.  Recall that for all $d$ the tensor product of two spin $j$ representations of SO($d+1$) contains an $SO(d+1)$-invariant singlet.  In our basis, most components of this singlet state have $k \neq 0$ for both particles, though one component has $k=0$ for both particles.  However, we can use this 2-particle singlet to build several distinct 4-particle singlets.  Let $|(12)(34)\rangle$ be the tensor product of the 2-particle singlet state formed from particles 1 and 2 with the 2-particle singlet state formed from particles 3 and 4, and similarly for $|(13)(24)\rangle$.  Then $|(12)(34)\rangle - |(13)(24)\rangle$ is a rotationally-invariant state in which every component has $k \neq 0$ for at least two of the four particles; the component with $k=0$ for all particles cancels between the two terms.  In this sense we can always find SO($D$)-invariant states in which $\sum_i k_i \ge N_4/2$, where $N_4$ is the largest integer less than or equal to $N$ which is divisible by $4$.  We can therefore find states where group-averaging converges whenever  $N_4 > 2d$, even when $\sigma$ is very close to zero.

\subsection*{Acknowledgements}
D.M. would like to thank Atsushi Higuchi and Steve Giddings for many discussions of group averaging in de Sitter space. The authors would also like to thank Atsushi Higuchi for sharing his unpublished notes \cite{AHnotes}. This work was supported in part by the US National Science Foundation under Grant No.~PHY05-55669, and by funds from the University of California.

\appendix

%%%%%%%%%%%%%%%%%%%%%%%%%%%%%%%%%%%%%%%%
\section{Symmetric tensor spherical harmonics of rank $r \le 2$} \label{app:STSHs}
%%%%%%%%%%%%%%%%%%%%%%%%%%%%%%%%%%%%%%%%

In this section we briefly review symmetric tensor spherical harmonics
(STSHs) on the $d$-sphere $S^d$. Tensor harmonics of general rank are
discussed in \cite{Higuchi:1986wu}.  We also provide a more detailed
discussion of spherical harmonics of rank $r \le 2$.

We use coordinates
$\Omega_d = (\theta_{d-1},\dots,\theta_1, \phi)$ on $S^d$.
The metric on $S^1$ is simply
$d\Omega_1^2 = d\phi^2$,
and on $S^d$ is
\eqn{ \label{eq:spheremetric}
  d\Omega_d^2
  &=& d\theta_{d-1}^2 + \sin^2\theta_{d-1} d\Omega_{d-1}^2 \nn \\
  &=& d\theta_{d-1}^2 + \sin^2\theta_{d-1}
  (d\theta_{d-2}^2 + \sin^2\theta_{d-2} ( \dots + \sin^2\theta_2
  (d\theta_1^2 + \sin^2\theta_1 \, d\phi^2 ) \dots ) ) .
}

%%%%%%%%%%%%%%%%%%%%%%%%%%%%%%%%%%%%%%%%
\subsection{Scalar STSHs} \label{app:scalar}
%%%%%%%%%%%%%%%%%%%%%%%%%%%%%%%%%%%%%%%%

Scalar spherical harmonics on $S^d$ are denoted
${}^{(d)}Y_{\vj}(\Omega_d)$. They are labeled by $d$ integers
$\vj = j_{d},\dots,j_1$ with
$j_{d} \ge j_{d-1} \ge \dots \ge j_2 \ge |j_1|$.
The harmonics satisfy
\eq{ \label{eq:scalarSH}
 \Laplaced \Yd_{\vj}(\Omega_d)
 = - j_d(j_d+d-1) \Yd_{\vj}(\Omega_d) \;,
}
where $\Laplaced$ is the scalar Laplace-Bertrami operator on $S^d$. For
$d=1$ this is$ {}^{(1)}\nabla^2 = \d_\phi^2$ and for $d>1$ we have
\eq{
  \Laplaced = \d_{\theta_{d-1}}^2
  + (d-1)(\cot\theta_{d-1})\d_{\theta_{d-1}}
  + \frac{\Laplacedd}{(\sin\theta_{d-1})^2}, \quad d > 1 \;.
}
The solutions to (\ref{eq:scalarSH}) are then
\eq{
  {}^{(1)}Y_{j_1}(\phi) := \frac{e^{-i j_1 \phi}}{\sqrt{2\pi}} \;,
  \quad
  {}^{(d)}Y_{\vj}(\Omega_d) =
  \left[ \prod_{n=2}^{d} {}^{(n)}\cY_{j_{n},j_{n-1}}(\theta_{n-1})
    \right]\, {}^{(1)}Y_{j_1}(\phi) , \quad d > 1 \;,
}
where
\eqn{
  {}^{(n)}\cY_{j k}(\theta) &:=&
  \left[\frac{(2j+n-1)\G(j+k+n-1)}{2\G(j-k+1)}\right]^{1/2}
  (\sin \theta)^{-(n-2)/2}
  P^{-(k+(n-2)/2)}_{j+(n-2)/2}(\cos \theta) \;.
}
Here $P^{-\mu}_\nu(x)$ is the associated Legendre function
\cite{GR}.
The function ${}^{(d)}\cY_{jk}(\theta)$ is normalized such that
\eq{
  \int_0^\pi d\theta (\sin\theta)^{d-1} \, {}^{(d)}\cY_{jk}(\theta)
  \overline{{}^{(d)}\cY_{j'k}(\theta)} = \delta_{jj'}
}
for $d>1$. As a result, the scalar harmonics are normalized with respect to
the inner product
\eq{
  ( Y_{\vj}, Y_{\vj'} )_{L^2}
  := \int d\Omega_d \, Y_{\vj}(\Omega_d) \overline{Y_{\vj'}(\Omega_d)}
  = \delta_{\vj \vj'}\;.
}

%%%%%%%%%%%%%%%%%%%%%%%%%%%%%%%%%%%%%%%%
\subsection{Vector STSHs} \label{app:vector}
%%%%%%%%%%%%%%%%%%%%%%%%%%%%%%%%%%%%%%%%

Following \cite{Higuchi:1986wu} we create vector STSHs on $S^d$ out of
scalar and vector STSHs on $S^{d-1}$. It is convenient to denote the
metric, covariant derivative, and Laplace operator on a sphere $S^n$
by ${}^{(n)}\gamma_{ab}$, ${}^{(n)}D_a$, and ${}^{(n)}\nabla^2$.
We also introduce the notation $\chi = \theta_{d-1}$.
Using our coordinates (\ref{eq:spheremetric}) the line element
on $S^d$ is
\eq{
  ds^2 = \gd_{ab} dx^a dx^b = d\chi^2 + (\sin\chi)^2\, \gdd_{ij}dx^i dx^j \;.
}
We let indices $a,b$ range over $S^d$ while indices $i,j$ will
have range only over the $S^{d-1}$.

Vector STSHs satisfy
\eq{ \label{eq:vectorHarmonicEquations}
  \Dd^a \,\Vd_a = 0 , \quad
  \Laplaced \, \Vd_a
  = \left[ - j(j+d-1) +1 \right] \Vd_a\
}
and depend on $d$ angular momenta $\vj$ that satisfy
$j_d \ge j_{d-1} \ge \dots \ge j_2 \ge |j_1|$.
As in the main text we set $j = j_{d}$ and $k = j_{d-1}$.
There are $d-1$ sets of orthogonal solutions to these equations,
one set each with non-vanishing components tangent to the sub-sphere
$S^2,S^3,\dots,S^D$. The sets are labeled $\Vd^{(2)},\Vd^{(3)},\dots,
\Vd^{(d)}$, and individual harmonics are also labeled by their
momenta $\vj$. These harmonics are given by
\eqn{\label{A10}
  \Vd^{(d;\vj)}_\chi(\Omega_d)
  &=& \left[\frac{k(k+d-2)}{(j+1)(j+d-2)}\right]^{1/2}
  (\sin\chi)^{-1} \, \cYd_{j k}(\chi) \,
  \Ydd_{\vk}(\Omega_{d-1})\;, \\
  \Vd^{(d;\vj)}_i(\Omega_d)
  &=&
  \left[\frac{1}{(j+1)(j+d-2)k(k+d-2)}\right]^{1/2} \nn \\ & &
  \times \sin\chi \left[
  \left(\d_\chi + (d-2)\cot\chi \right) \cYd_{jk}(\chi)
  \right] \, \Ddd_i \, \Ydd_{\vk}(\Omega_{d-1})\;, \\
  \Vd^{(\alpha;\vj)}_\chi(\Omega_d) &=& 0, \quad \alpha \neq d \\
  \Vd^{(\alpha;\vj)}_i(\Omega_d) &=&
  (\sin\chi) \, \cYd_{jk}(\chi) \, \Vdd_i^{(\alpha;\vk)}(\Omega_{d-1})\;,
  \quad \alpha \neq d \;.
}
Since \eqref{A10} vanishes for $k=0$, we see that non-trivial harmonics exist only for $j \ge k \ge 1$.
The vector harmonics are normalized with respect to the inner
product
\eq{
  ( \Vd^{(1)} , \Vd^{(2)} )_V
  := \int d\Omega_d \,\Vd^{(1)}_{a} \,
  \overline{\Vd^{(2)}}^{a} = \delta^{(1)(2)}.
}

%%%%%%%%%%%%%%%%%%%%%%%%%%%%%%%%%%%%%%%%
\subsection{Residual Gauge Transformations}
\label{residual}
%%%%%%%%%%%%%%%%%%%%%%%%%%%%%%%%%%%%%%%%

This appendix deals with certain issues related to the residual
gauge transformations noted in section \ref{sec:vector} preserving the gauge condition $\nabla^\mu A_\mu =0$.  Such transformations
take the form $A_\mu \rightarrow A_\mu + \ell \nabla_\mu \varphi_0$,
where $\varphi_0$ satisfies the Klein-Gordon equation with $\sigma=0$.
These transformations can be used to entirely eliminate the modes
$A_\mu^{(D;\vj)}$, and also have implications for the modes
$A_\mu^{(d;\vj)}$.  We address each in turn.

\bigskip

{\bf The modes $A_\mu^{(D;\vj)}$:}

\bigskip

The above residual gauge transformations can be used to eliminate the
modes $A_\mu^{(D;\vj)}$. The point is simply that
$A_\mu^{(D;\vj)} \propto  \nabla_\mu \varphi_{0 ,\vj}$, so that these
modes are pure gauge.

We begin by introducing some new tools. Note that the following Casimir-raising and
lowering operators
\eqn{ \label{eq:sigmaOperators}
   \hat{L}_{\sigma} &:=& i (\cosh t \d_t - \sigma \sinh t), \nn \\
  \hat{R}_{\sigma} &:=& i (\cosh t \d_t + (\sigma+d) \sinh t),
}
act on $\chY_{\sigma j}(t)$ as
\eqn{
  \hat{L}_{\sigma} \chY_{\sigma j}(t) &=&
  \left[(j-\sigma)(j+\sigma-1+d)\right]^{1/2} \chY_{\sigma-1,j}(t), \nn \\
  \hat{R}_{\sigma} \chY_{\sigma j}(t) &=&
  \left[(j-\sigma-1)(j+\sigma+d)\right]^{1/2} \chY_{\sigma+1,j}(t).
}

The modes $A^{(D;\vj)}_\mu$ are explicitly
\eqn{
  A^{(D;\vj)}_t &=&
  N^{(t)}_{\sigma_1 j} (\cosh t)^{-1} \chY_{\sigma_1 j}(t)
  \Yd_{\vj}(\Omega_d) \;,\\
  A^{(D;\vj)}_a &=&
  - \frac{N^{(t)}_{\sigma_1 j}}{j(j+d-1)}(\cosh t)
  \left[
    (\d_t + (d-1) \tanh t) \chY_{\sigma_1 j}(t) \right]
  \Dd_a \Yd_{\vj}(\Omega_d) \;,
}
where $N^{(t)}_{\sigma_1 j}$ is a normalization constant whose
value is not important here. By inspection we see
\eqn{ \label{eq:dtphi}
  \nabla_t \varphi_{0, \vj}
  &=&  \d_t \chY_{0,j}(t) \Yd_{\vj}(\Omega_d) \nn \\
  &=&  \left[ -i (\cosh t)^{-1} \hat{L}_0 \chY_{0,j}(t)\right]
  \Yd_{\vj}(\Omega_d) \nn \\
  &=& \frac{- i}{ \left[j(j+d-1)\right]^{1/2}} (\cosh t)^{-1}
  \chY_{-1,j}(t) {}^{(d)}\Yd_{\vj}(\Omega_d) \nn \\
  &=& \frac{-i}{\left[j(j+d-1)\right]^{1/2}N^{(t)}_{\sigma_1 j}} A_t^{(t;\vj)} ,
}
and
\eqn{
  A_a^{(t;\vj)}
  &=& i \frac{N^{(t)}_{\sigma_1 j}}{j(j+d-1)}
  \left[\hat{R}_{-1}\chY_{-1,j}(t) \right]
  \Dd_a \Yd_{\vj}(\Omega_d) \nn \\
  &=&  i N^{(t)}_{\sigma_1 j} [j(j+d-1)]^{1/2} \chY_{0,j}(t)
  \Dd_a \Yd_{\vj}(\Omega_d) \nn \\
  &=&  i N^{(t)}_{\sigma_1 j} [j(j+d-1)]^{1/2} \,\nabla_a \varphi_{0,\vj} .
}
This proves  that $ A^{(D;\vj)}_\mu \propto \nabla_\mu \varphi_{0, \vj}$.

\bigskip

{\bf The modes $A_\mu^{(d;\vj)}$:}

\bigskip

Although the modes $A_\mu^{(d;\vj)}$ should define an irreducible
representation of the de Sitter group, the action of $\pounds_B$
on $A_\mu^{(d;\vj)}$ is not a linear combination of such modes. The
issue is simply that $\pounds_B A_\mu^{(d;\vj)}$ also contains a pure
gauge term which must be removed; i.e, the term
$c \ell \nabla_\mu \varphi_{0}$ in $\delta_B$ (\ref{dbeq}) is non-trivial.

We begin by examining the $t$ and $\chi$ components in detail:
\eqn{
  A^{(d;\vj)}_t    &=& 0 \;,\nn\\
  A^{(d;\vj)}_\chi &=&
  \left[\frac{k(k+d-2)}{(j+1)(j+d-2)}\right]^{1/2}
  \frac{\cosh t}{\sin\chi} \chY_{-1,j}(t) \cYd_{jk}(\chi)
  \Ydd_{\vk}(\Omega_{d-1})\;.
}
First consider the boost action on $A_t^{(d;\vj)}$; because
$A_t^{(d;\vj)}=0$ it is easy to extract the pure gauge
contribution:
\eqn{ \label{eq:deltaB1}
  \delta_B A_t^{(d;\vj)}
  &=& - i \frac{\sin\chi}{\cosh^2 t}A_\chi^{(d;\vj)}
   -  c \ell \nabla_t \varphi_{0,\vj} = 0\;.
}
Next we use the $\sigma$ raising and lowering operators
(\ref{eq:sigmaOperators}) to note
\eq{
  \hat{R}_{\sigma = -1} \hat{L}_{\sigma = 0} \varphi_{0,j} =
  - (\cosh t \,\d_t + (d-1)\sinh t )(\cosh t \,\d_t) \varphi_{0,\vj}
  = j(j+d-1) \varphi_{0,\vj}
}
(this can also be seen from (\ref{eq:scalarBox})). We can
now use this equation to invert (\ref{eq:deltaB1}), namely
\eq{
  i
  (\cosh t \,\d_t + (d-1) \sinh t) \frac{\sin\chi}{\cosh t} A_\chi^{(d;\vj)}
  = c j(j+d-1) \varphi_{0,\vj}\;,
}
from which we find
\eq{
   c \ell \nabla_\mu \varphi_{0,\vj}
  = \frac{i}{j(j+d-1)} \nabla_\mu
  \left[\left(\cosh t \,\d_t + (d-1)\sinh t\right)
    \frac{\sin\chi}{\cosh t} A_\chi^{(d;\vj)} \right]\;.
}
We now have an explicit expression for the pure gauge contribution
to the $A_\mu^{(d;\vj)}$ modes.

%%%%%%%%%%%%%%%%%%%%%%%%%%%%%%%%%%%%%%%%
\subsection{Tensor STSHs} \label{app:tensor}
%%%%%%%%%%%%%%%%%%%%%%%%%%%%%%%%%%%%%%%%

Finally, we turn to rank $2$ STSHs on the sphere $S^d$.
These harmonics $\Td_{a b}$ satisfy
\eq{ \label{eq:tensorHarmonicEquations}
  \gd^{a b} \, \Td_{a b} = 0 , \quad
  \Dd^a \, \Td_{a b} = 0, \quad
  \Laplaced \,\Td_{a b} = \left[ - j (j + d-1) + 2 \right] \Td_{a b} .
}
The solutions depend on $d$ angular momenta $\vj$ which satisfy
$j_d \ge j_{d-1} \ge \dots \ge j_2 \ge |j_1| \ge 2$. The
$(d+1)(d-2)/2$ independent sets of solutions to
(\ref{eq:tensorHarmonicEquations})
are have non-vanishing components tangent to the sub-spheres
$S^2,S^3,\dots,S^d$. The harmonics are thus labeled
$\Td^{(\alpha\beta;\vj)}_{ab}$ where $\alpha,\beta$ run over the
sub-spheres $3,\dots,d$. Explicit expressions for the harmonics are
\eqn{
  T^{(dd;\vj)}_{\chi\chi}
  &=& N^{(dd)}_{jk} (\sin\chi)^{-2} \, \cYd_{jk}(\chi)
  \,\Ydd_{\vk}(\Omega_{d-1})\;, \nn\\
  T^{(dd;\vj)}_{\chi i}
  &=& \frac{N^{(dd)}_{jk}}{k(k+d-2)}
  \left[\d_\chi + (d-2)\cot\chi \right] \cYd_{k\ell}(\chi) \,
  \Ddd_i \Ydd_{\vk}(\Omega_{d-1})\;, \nn\\
%  T^{(dd;\vj)}_{i i'}
%  &=& \dots \nn\\
  T^{(d \alpha; \vj)}_{\chi \chi} &=& 0 , \quad \alpha \neq d\;,\nn\\
  T^{(d \alpha; \vj)}_{\chi i}
  &=& N_{jk}^{(d)} \cYd_{jk}(\chi)\,
  \Vdd^{(\alpha;\vk)}_i(\Omega_{d-1}),  \quad \alpha \neq d\;, \nn\\
  T^{(d \alpha; \vj)}_{i i'}
  &=& \frac{2 N_{jk}^{(d)}}{\left[k(k+d-2)-2\right]} (\sin \chi)^2
  \left[ \d_\chi + (d-1)\cot \chi \right] \cYd_{jk}(\chi)\, \Ddd_{(i}
  \Vdd^{(\alpha;\vk)}_{i')}(\Omega_{d-1}), \quad \alpha \neq d\;, \nn\\
  T^{(\alpha \beta; \vj)}_{\chi \chi}
  &=& T^{(\alpha \beta; \vj)}_{\chi i} = 0 , \quad \alpha, \beta \neq d\;,\nn\\
  T^{(\alpha\beta;\vj)}_{i i'}
  &=& (\sin\chi)^2\,\cYd_{jk}(\chi)
  \,\Td^{(alpha\beta;\vk)}_{ii'}(\Omega_{d-1}) ,\quad
  \alpha, \beta \neq d\;.
}
The normalization constants are
\eq{
  N^{(d)}_{jk}
  = \left[ \frac{(k-1)(k+d-1)}{2 j (j+d-1)} \right]^{1/2},
  \quad
  N^{(dd)}_{jk}
  = \left[ \frac{ (d-2)(k-1)k(k+d-2)(k+d-1)}
  { (d-1) j (j+1)(j+d-2)(j+d-1)} \right]^{1/2} \;.
}
Tensor harmonics are normalized with respect to the inner product
\eq{
  ( T^{(1)} , T^{(2)} )_T
  := \int d\Omega_d \,T^{(1)}_{ab} \, \overline{T^{(2)}}^{ab} = \delta^{(1)(2)}.
}

%%%%%%%%%%%%%%%%%%%%%%%%%%%%%%%%%%%%%%%%%%%%%%%%%%%%%%%%%%%%%%%%%
\section{Finite boost matrix elements} \label{app:integral}
%%%%%%%%%%%%%%%%%%%%%%%%%%%%%%%%%%%%%%%%%%%%%%%%%%%%%%%%%%%%%%%%%

In this appendix we compute matrix elements
$B_{\vj\vm}(\lambda) = \KG{B(\lambda) \varphi_{\sigma \vj}}{\varphi_{\sigma \vm}}$.
This reduces to computing the integral
\eqn{
  B_{j m k}(\l)
  = - 2 i \left[\d_t \overline{\chY}_{\sigma m}(t)\right]
  \int_0^\pi d\chi (\sin\chi)^{d-1}
  \,\chY_{\sigma j}(t') \cYd_{j k}(\chi')
   \overline{\cYd}_{m k}(\chi) \bigg|_{t=0} \;,
   \label{eq:int}
}
where $\chY_{\sigma j}(t)$ and $\cYd_{j k}(\chi)$ are defined in terms
of Legendre functions in (\ref{eq:chY}) and (\ref{eq:scalarSH})
respectively, and $t',\chi'$ are defined in (\ref{eq:coordinateChange}).
The Legendre function may be defined in terms of the Gauss hypergeometric
function via
\eq{ \label{eq:LegendreDef}
  P^{-m}_{n}(z) := \frac{1}{\G(1+m)} \left[\frac{z-1}{z+1}\right]^{m/2}
  \2F1{-n}{1+n}{1+m}{\frac{1-z}{2}} .
}
After some standard manipulations of the hypergeometric function
we may write $\chY_{\sigma j}(t)$ and $\cYd_{j k}(\chi)$ as
\eqn{
  \chY(t)_{\s j} &=&
  \wN_{d\s j} (\cosh t)^{j} \nn \\ & & \times
  \sum_{f=0,1} C_f (\sinh t)^f
  \2F1{\frac{j-\s+f}{2}}{\frac{j+\s+d}{2}}{\half+f}{-\sinh^2 t} , \\
  \cYd(\chi)_{jk}
  &=& \wM_{d j k} (\sin \chi)^{k} (\cos \chi)^{j-k}
  \2F1{\frac{k-j}{2}}{\frac{k-j+1}{2}}{k+\frac{d}{2}}{-\tan^2 \chi} ,
}
where the constants $\wN_{d \s j}$, $\wM_{d j k}$
and $C_f$ are
\eqn{
  \label{eq:wN}
  \wN_{d \s j} &=&
  \frac{1}{2^{j+(d-1)/2}\G\left(j+\frac{d+1}{2}\right)}
  \left[\frac{\GG{j-\s,j+\s+d}}{2}\right]^{1/2} , \\
  \label{eq:wM}
  \wM_{d j k} &=&
  \frac{1}{2^{k+(d-2)/2}\G\left(k+\frac{d}{2}\right)}
  \left[\left(j + \frac{d-1}{2}\right)
    \GGG{j+k+d-1}{j-k+1}\right]^{1/2} , \\
  \label{eq:Cf}
  C_f &=& (-i)^f
  \GGG{\half-f, j+\frac{d+1}{2}}{\frac{j-\s+1-f}{2}, \frac{j+\s+d+1-f}{2}} .
}
Here $\Gamma[\dots]$ denotes products (and quotients) of gamma functions.
Utilising these expressions and noting that
\eq{
  \left[ \d_t  \overline{\chY}_{\s m}(t) \right]_{t = 0}
  = 2 i \sqrt{\pi} \wN_{d \s m}
  \GGG{m+\frac{d+1}{2}}{\frac{m-\s}{2},\frac{m+\s+d}{2}}
}
the boost matrix elements (\ref{eq:int}) become
\eqn{
  B_{j m k}(\l)
  &=& 4 \sqrt{\pi} \wN_{d \s j} \wN_{d \s m} \wM_{d j k} \wM_{d m k}
  \GGG{m+\frac{d+1}{2}}{\frac{m-\s}{2},\frac{m+\s+d}{2}} I
}
where $I$ is the integral
\eqn{
  \label{eq:I}
  I &=& \sum_{f=0,1} C_f
  \int_0^\pi d\chi \Bigg\{(\sin\chi)^{d-1+k}(\cos\chi)^{m-k}(\sin\chi')^k
  (\cos\chi')^{j-k}(\cosh t')^j(\sinh t')^f
  \nn \\ & & \ph{\sum_{f=0,1} C_f\int_0^\pi d\chi \Bigg\{}
  \times \2F1{\frac{k-m}{2}}{\frac{k-m+1}{2}}{k+\frac{d}{2}}{-\tan^2 \chi}
  \nn \\ & & \ph{\sum_{f=0,1} C_f\int_0^\pi d\chi \Bigg\{}
  \times \2F1{\frac{k-j}{2}}{\frac{k-j+1}{2}}{k+\frac{d}{2}}{-\tan^2 \chi'}
  \nn \\ & & \ph{\sum_{f=0,1} C_f\int_0^\pi d\chi \Bigg\{}
  \times
  \2F1{\frac{j-\s+f}{2}}{\frac{j+\s+d}{2}}{\half+f}{-\sinh^2 t'} \Bigg\} .
}

Let us now focus on $I$. First note that at $t = 0$ the boosted
coordinates satisfy (see (\ref{eq:coordinateChange}))
\eqn{
  \sinh t' &=& \sinh \l \cos \chi, \quad
  \sin\chi' \cosh t' = \sin \chi, \quad (t = 0)\\
  \cos\chi' \cosh t' &=& \cosh \l \cos \chi, \quad
  \tan^2 \chi' = \cosh^{-2}\l \tan^2\chi, \quad (t = 0).
}
We may use these relations to eliminate the boosted coordinates
from (\ref{eq:I}):
\eqn{
  I &=& (\cosh\l)^{j-k}
  \sum_{f=0,1} C_f (\sinh \l)^f \nn \\ & & \times
  \int_0^\pi d\chi \Bigg\{(\sin\chi)^{d-1+2k}(\cos\chi)^{j+m+f-2k}
  \nn \\ & & \ph{\times \int_0^\pi d\chi \Bigg\{}
  \times \2F1{\frac{k-m}{2}}{\frac{k-m+1}{2}}{k+\frac{d}{2}}{-\tan^2 \chi}
  \nn \\ & & \ph{\times \int_0^\pi d\chi \Bigg\{}
  \times
  \2F1{\frac{k-j}{2}}{\frac{k-j+1}{2}}{k+\frac{d}{2}}{- \cosh^{-2}\l \tan^2 \chi}
  \nn \\ & & \ph{\times \int_0^\pi d\chi \Bigg\{}
  \times
  \2F1{\frac{j-\s+f}{2}}{\frac{j+\s+d}{2}}{\half+f}{-\sinh^2 \l \cos^2\chi}
  \Bigg\} .
}
We can now expand the first two hypergeometric functions into their
power series representations, i.e.,
\eq{
  \2F1{\frac{k-m}{2}}{\frac{k-m+1}{2}}{k+\frac{d}{2}}{-\tan^2 \chi}
  = \sum_{a=0}^{[(m-k)/2]} \psi_a (-\tan^2\chi)^a,
}
where with condensed notation we take $\psi_a$ to represent the
coefficient of the power series
\eq{
  \psi_a :=
  \frac{\left(\frac{k-m}{2}\right)_a \left(\frac{k-m+1}{2}\right)_a}
  {\left(k + \frac{d}{2}\right)_a \G(1+a)} ,
}
and $(a)_n := \G(a+n)/\G(a)$ is the Pochhammer symbol. With this
$I$ becomes
\eqn{
  I &=& (\cosh\l)^{j-k}
  \sum_{f=0,1} \sum_{a=0}^{[(m-k)/2]} \sum_{b=0}^{[(j-k)/2]}
  \Bigg\{ C_f \psi_a \psi_b (-1)^{a+b} (\cosh \l)^{-2b} (\sinh \l)^f
  \nn \\ & & \times
  \int_0^\pi d\chi (\sin\chi)^{d-1+2k+2a+2b}(\cos\chi)^{j+m+f-2k-2a-2b}
  \nn \\ & & \ph{\times \int_0^\pi d\chi }
  \times
  \2F1{\frac{j-\s+f}{2}}{\frac{j+\s+d}{2}}{\half+f}{-\sinh^2 \l \cos^2\chi}
  \Bigg\} .
  \label{eq:I2}
}
Because the integral is over the region $0$ to $\pi$, the integral
is non-vanishing only if the integrand has an even number of $\cos\chi$.
Thus of the two terms $f = 0,1$, only one survives: $f=0 (1)$ if
$j+m$ is even (odd).

The integral (\ref{eq:I2}) can now be computed using the formula
\eq{
  \int_0^1 dy\; y^{c-1} (1-y)^{d-c-1}
  {}_A F_B\left[\vec{a}\,;\,\vec{b}\,;\, z y\right]
  = \GGG{c, d-c}{d}
  {}_{A+1} F_{B+1}\left[\vec a\,,\, c\,;\,\vec b\,,\, d\,; z\right] ,
}
where $\vec a$ and $\vec b$ are lists of parameters of length
$A$ and $B$ respectively. Thus $I$ is
\eqn{
  I &=& \frac{C_f}{\G\left(\frac{j+m+f+d+1}{2}\right)}
  (\cosh\l)^{j-k} (\sinh \l)^f \nn \\ & & \times
  \sum_{a=0}^{[(m-k)/2]} \sum_{b=0}^{[(j-k)/2]}
  \Bigg\{
  \psi_a \psi_b (-1)^{a+b}
  \GG{k+\frac{d}{2}+a+b,\frac{j+m+f+1}{2}-k-a-b}
  (\cosh \l)^{-2b}
  \nn \\ & & \ph{\times \sum_{a=0}^{[(m-k)/2]} \sum_{b=0}^{[(j-k)/2]}
  \Bigg\{} \times
  \3F2{\frac{j-\sigma+f}{2}}{\frac{j+\sigma+d+f}{2}}{\frac{j+m+f+1}{2}-k-a-b}
  {\half+f}{\frac{j+m+d+1+f}{2}}{-\sinh^2 \lambda} \Bigg\} .\nn \\
}

Using this result we can now write the boost matrix elements as
\eqn{
  B_{j m k}(\l) &=&
  \mathcal{A}_{j m k}^\s (\cosh \l)^{j-k} (i\sinh \l)^f
  \Bigg\{ \sum_{a=0}^{[(m-k)/2]} \sum_{b=0}^{[(j-k)/2]}
  \mathcal{B}_{j m k}^{a b}
  (\cosh \l)^{-2 b} \nn \\ & &
  \times
  \3F2{\frac{j-\sigma+f}{2}}{\frac{j+\sigma+d+f}{2}}{\frac{j+m+f+1}{2}-k-a-b}
  {\half+f}{\frac{j+m+d+1+f}{2}}{-\sinh^2 \lambda} \Bigg\}
}
where explicitly
\eqn{
  \label{eq:curlyADef}
  \mathcal{A}_{j m k}^\s &=&
  \frac{1}{2^{2k+d-2-f}\G\left(\frac{j+m+f+d+1}{2}\right)}\nn \\ & &\times
  \left\{
    \GGG
    {\frac{m-\s+1}{2},\frac{m+\s+d+1}{2},\frac{j-\s+f}{2},\frac{j+\s+d+f}{2},
      j+k+d-1,m+k+d-1}
    {\frac{m-\s}{2},\frac{m+\s+d}{2},\frac{j-\s+1-f}{2},\frac{j+\s+d+1-f}{2},
      1+j-k,1+m-k}
  \right\}^{1/2} , \nn \\
  & & \\
  \mathcal{B}_{j m k}^{a b}
  &=& \left(\frac{-1}{4}\right)^{a+b}
  ( k-m )_{2a}\, ( k-j )_{2b}\,
  \GGG{k+\frac{d}{2}+a+b,\frac{j+m+f+1}{2}-k-a-b}
  {k + \frac{d}{2}+a,1+a,k + \frac{d}{2}+b,1+b} .
  \label{eq:curlyBDef}
}

%%%%%%%%%%%%%%%%%%%%%%%%%%%%%%%%%%%%%%%%%%%%%%%%%%%%%%%%%%%%%%%%%

%%%%%%%%%%%%%%%%%%%%%%%%%%%%%%%%%%%%%%%%

%%%%%%%%%%%%%%%%%%%%%%%%%%%%%%%%%%%%%%%%
\end{document}